\documentclass[copyright,creativecommons]{eptcs}
\providecommand{\event}{FIT 2010}
\usepackage{breakurl} 

\usepackage{amssymb,xspace,epsfig,color}
\usepackage[latin1]{inputenc}
\usepackage{listings,multibib,url,wrapfig}
\usepackage{multicol,amsmath}
\usepackage{float}
\usepackage{subfig}

\graphicspath{{images/}}

\makeatletter
\newbox\sf@box
\newenvironment{SubFloat}[2][]{%
\def\sf@one{#1}%
\def\sf@two{#2}%
\setbox\sf@box\hbox
\bgroup
}{%
\egroup
\ifx\@empty\sf@two\@empty\relax
\def\sf@two{\@empty}
\fi
\ifx\@empty\sf@one\@empty\relax
\subfloat[\sf@two]{\box\sf@box}%
\else
\subfloat[\sf@one][\sf@two]{\box\sf@box}%
\fi
}

\lstdefinelanguage{Esterel}
{morekeywords={abort, and, await, call, case, combine, constant, do,
    each, emit, else, elseif, end, every, exec, exit, function, halt,
    handle, if, immediate, in, input, inputoutput, loop, module, not,
    nothing, or, output, pause, positive, pre, present, procedure,
    relation, repeat, return, run, sensor, signal, suspend, sustain,
    task, then, times, trap, type, var, weak, when, with},
  morecomment=[l]{\%}
}

\lstdefinelanguage{KEP}
{morekeywords={ ABORT, ADD, ADDC, AWAIT, AWAITI, AWAITN, CAWAIT,
    CAWAITE, CMP, EMIT, EMITD, EMITR, GOTO, HALT, INPUT, JOIN, JW,
    LOAD, MUL, NOTHING, OUTPUT, PAUSE, PAR, PARE, PRE, PRIO, PRESENT,
    SIGNAL, SUSPEND, SUSPENDI, SUSTAIN, SUSTAIND, SUSTAINR, WABORT, WABORTI,
  },
  morecomment=[l]{//},
  morecomment=[s]{/*}{*/},
  morecomment=[l]{\%},
  morestring=[b]",
  numbers=none,
}

\newcommand{\ie}{\emph{i.\,e.}\xspace}

\newcommand{\eg}{\emph{e.\,g.}\xspace}

\newcommand{\etal}{\emph{et al.}\xspace}

\newcommand{\emit}{\textsf{emit}\xspace}
\newcommand{\present}{\textsf{present}\xspace}
\newcommand{\pause}{\textsf{pause}\xspace}
\newcommand{\halt}{\textsf{halt}\xspace}
\newcommand{\nothing}{\textsf{nothing}\xspace}

\newcommand{\goto}{\textsf{goto}\xspace}
\newcommand{\wabort}{\textsf{wabort}\xspace}
\newcommand{\fork}{\textsf{fork}\xspace}
\newcommand{\join}{\textsf{join}\xspace}
\newcommand{\PAR}{\textsf{PAR}\xspace}
\newcommand{\PARE}{\textsf{PARE}\xspace}
\newcommand{\JOIN}{\textsf{JOIN}\xspace}

\newcommand{\Nat}{\ensuremath{{\mathbb N}}\xspace}
\newcommand{\cNat}{\ensuremath{{\mathbb N}_{\infty}}\xspace}

\newcommand{\df}{=_{\scriptstyle{df}}}
\newcommand{\miff}{\mbox{\it iff\/}}
 
\mathchardef\land="2026

\newcommand{\sem}[1]{[ \! [ #1 ] \! ]}
\newcommand{\set}[2]    {\ensuremath{\{\,#1\,\mid\,#2\,\}}}

\newcommand{\length}[1]{\ensuremath{|#1|}\xspace}

\newcommand{\sigs}{\ensuremath{\mathbb{S}}\xspace}
\newcommand{\vars}{\ensuremath{\mathbb{V}}\xspace}
\newcommand{\labs}{\ensuremath{\mathbb{L}}\xspace}
\newcommand{\states}{\ensuremath{\mathbb{M}}\xspace}

\newcommand{\imp}{\ensuremath{\supset}\xspace}
\newcommand{\true}{\mbox{\textit{true}}\xspace}
\newcommand{\false}{\mbox{\textit{false}}\xspace}

\newcommand{\delm}{\ensuremath{{\circ}}\xspace}

\newcommand{\maxd}{\ensuremath{\textit{max}}\xspace}
\newcommand{\mind}{\ensuremath{\textit{min}}\xspace}

\mathcode`:="003A               
\mathchardef\rcol="303A         

\newcommand{\sbound}[1]{\ensuremath{\textit{Bnd}(#1)\xspace}}
\newcommand{\nset}[1]{\ensuremath{\underline{#1}}}
\newcommand{\plus}{\ensuremath{+}}

\newcommand{\start}{\textit{out}\xspace}
\newcommand{\wait}{\mbox{\textit{in}}\xspace}

\newtheorem{definition}{Definition}[section]
\newtheorem{proposition}[definition]{Proposition}

\newcommand{\qed}{}
\newcommand{\Proof}{\par\medskip\noindent{\bf Proof:\ }}

\title{%
  An Algebra of Synchronous Scheduling Interfaces 
}

\author{Michael Mendler
\institute{Faculty of Information Systems and Applied Computer Sciences \\ 
             Bamberg University}
\email{michael.mendler@uni-bamberg.de}
}

\def\authorrunning{Michael Mendler}
\def\titlerunning{An Algebra of Synchronous Scheduling Interfaces}

\begin{document}

\maketitle

\begin{abstract}

  In this paper we propose an algebra of synchronous scheduling
  interfaces which combines the expressiveness of Boolean algebra for
  logical and functional behaviour with the min-max-plus arithmetic
  for quantifying the non-functional aspects of synchronous
  interfaces. The interface theory arises from 
  a realisability interpretation of intuitionistic modal logic (also
  known as Curry-Howard-Isomorphism or propositions-as-types
  principle). The resulting algebra of interface types aims to provide
  a general setting for specifying type-directed and compositional
  analyses of worst-case scheduling bounds. It covers synchronous
  control flow under concurrent, multi-processing or
  multi-threading execution and permits precise statements about
  exactness and coverage of the analyses supporting a variety of
  abstractions. The paper illustrates the expressiveness of the
  algebra by way of some examples taken from network flow problems,
  shortest-path, task scheduling and worst-case reaction times in
  synchronous programming.

\end{abstract}

\section{Introduction}

The algebra discussed in this paper aims at the specification of
behavioural interfaces under the execution model of synchronous
programming. Such interfaces abstract externally observable Boolean
controls for components activated under the regime of a global
synchronous scheduler familiar from data-flow oriented languages such
as Lustre~\cite{Halbwachs05}, Signal~\cite{GuernicGBM91}, Lucid
Synchrone~\cite{lucy:manual06}, or imperative control-flow oriented
languages such as Statecharts~\cite{HarPnuPruShe87,PnueliS91},
Esterel~\cite{BerryG92} and Quartz~\cite{Schneider02}.  In this model
computations are coordinated under one or more global system clocks,
which may be physical or logical. They divide physical time into a
sequence of discrete \emph{ticks}, or \emph{instants}.  During each
instant the synchronous components interact using \textit{broadcast
  signals}, which can have one of two statuses, \textit{present} or
\textit{absent}. These signal statuses evolve monotonically as they
are propagated through the system, generating the emission or
inhibition of further signals and computations. Under the
\textit{synchrony hypothesis}~\cite{Halbwachs98} it is assumed that at
each instant, outputs are synchronous with the inputs. In other words,
computations take place instantaneously and appear to happen at each
tick ``all at once.''

The synchrony hypothesis conveniently abstracts internal, possibly
distributed computations into atomic reactions, making signals appear
almost like Boolean variables and (stateful) interfaces almost like
Mealy automata with Boolean labels.  Unfortunately, this abstraction
is not perfect, so that Boolean algebra is insufficient.  First, it is
well-known \cite{Hui91,mvm01-tocl} that classical two-valued Boolean
analysis is inadequate to handle the causality and compositionality
problems associated with the synchrony hypothesis adequately. E.g.,
Boolean algebra by itself cannot guarantee there are no races between
signal presence and absence, thus guaranteeing unique convergence
after a finite number of signal propagation steps. Some form of causality
information needs to be preserved. Secondly, quite practically, in
many applications we want to compute non-Boolean information about
otherwise ``instantaneous'' control signals, such as latency or
worst-case reaction times, maximal throughput, earliest deadlines, or
other quantitative information about the scheduling process.  This
provides one way to motivate the work reported here, viz.\ the search
for a fully abstract synchronisation algebra as an economic
refinement of classical Boolean algebra in situations where
Booleans are subject to synchronous schedules and quantitative
resource consumption.

Another motivation may be drawn from the arithmetical point of view.
One of the challenges in quantitative resource analysis is the clever
interchange (distribution) of $\maxd$, $\mind$ and $+$. For instance,
consider the analysis of worst-case reaction times (WCRT).  In its
simplest form, given a weighted dependency graph, the WCRT is the
maximum of all sums of paths delays, an expression of the form
$\maxd(\sum_{i\in p_1} d_{i1}, \sum_{i\in p_2} d_{i2}, \ldots,
\sum_{i\in p_n} d_{in})$ where $p_j$ are execution paths of the system
and $d_{ij}$ the delay of path segment $i$ in path $p_j$. As it
happens, the number $n$ of paths is exponential in the number of
elementary nodes of a system. Practicable WCRT analyses therefore
reduce the \textit{max-of-sums} to the polynomial complexity of
\textit{sum-of-maxes} (dynamic programming on dependency graphs) 
employing various forms of
dependency abstraction.  For illustration, imagine two alternative
path segments of length $d_1$, $e_1$ sequentially followed by two
alternative path segments of length $d_2$, $e_2$, respectively.  The
distribution $\maxd(d_1 + d_2, d_1 + e_2, e_1 + d_2, e_1 + e_2) =
\maxd(d_1, e_1) + \maxd(d_2, e_2)$ for efficiently calculating the
longest possible path, is exact only if we have a full set of path
combinations. In general, there will be dependencies ruling out
certain paths, in which case sum-of-maxes obtains but conservative
over-approximations. E.g., assume the combination of $d_1$ with $e_2$
is infeasible. Then, the sum-of-maxes is not exact since $\maxd(d_1,
e_1) + \maxd(d_2, e_2) \geq \maxd(d_1 + d_2, e_1 + d_2, e_1 + e_2)$.
On the other hand, knowing the infeasibility of $d_1 + e_2$ we would
rather compute $\maxd(d_1 + d_2, e_1 + \maxd(d_2, e_2)) = \maxd(d_1 +
d_2, e_1 + d_2, e_1 + e_2)$ which eliminates one addition and thus is
both exact \textit{and} more efficient than the full conservative
max-of-sums. The same applies to min-plus problems such as shortest
path or network flow. In the former, the efficient
\textit{sum-of-mins} is an under-approximation of the exact
\textit{min-of-sums} on all feasible paths. For network flow the
arithmetic is complicated further by the fact that min/max do not
distribute over $+$, i.e., $\mind(d, e_1 + e_2) \neq \mind(d, e_1) +
\mind(d, e_2)$ which obstructs simple linear programming techniques.

The art of scheduling analysis consists in finding a judicious
trade-off between merging paths early in order to aggregate
data on the one hand, and refining dependency paths by case analysis
for the sake of exactness, on the other hand.  A scheduling algebra
for practicable algorithms must be able to express and control this
trade-off.  In this paper we present an interface theory which
achieves this by coupling resource weights $d$ with logic formulas
$\phi$. A pair $d \rcol \phi$ specifies the semantic meaning of $d$
within the control-flow of a program module.  Logical operations on
the formulas then go hand-in-hand with arithmetic operations on
resources.  E.g., suppose a schedule activates control points $X$ and
$Y$ with a cost of $d_1$ and $d_2$, respectively, expressed $d_1 \rcol
\delm X \wedge d_2 \rcol \delm Y$. If the threads are resource
concurrent then both controls are jointly active within the maximum,
i.e., $\maxd(d_1, d_2) \rcol \delm(X \wedge Y)$.  If we are only
concerned whether one of the controls is reached, then we take
the minimum $\mind(d_1, d_2) \rcol \delm(X \oplus Y)$.  If activations
of $X$ and $Y$ requires interleaving of resources, then we must use
addition $d_1 + d_2 \rcol \delm(X \otimes Y)$. 

Our interface theory combines min-max-plus algebra $(\cNat, \mind,
\maxd, +, 0, -\infty, +\infty)$, see e.g.~\cite{BaccelliCOQ:SyncNLin},
with a refinement of Boolean algebra to reason about logical
control-flow. It features two conjunctions $\wedge$, $\otimes$ to
distinguish concurrent from multi-threading parallelism, two
disjunctions $\vee$, $\oplus$ to separate external from internal
scheduling choices, respectively. A consequence of its constructive
nature, our algebra replaces classical negation by a weaker and more
expressive \textit{pseudo-complement} for which
$\overline{\overline{x}} = x$ and $x + \overline{x} = 1$ are no longer
tautologies. This turns Boolean into a so-called Heyting algebra.  The
work presented here is an extension and adaptation of our earlier work
on \textit{propositional stabilisation theory}~\cite{mvm00:IGPL} which
has been developed to provide a semantic foundation for combinational
timing analyses.

The plan for the paper is as follows: To start with,
Sec.~\ref{sec:execution-schedules} lays out the syntactic and
semantical ground-work for our interface type theory which is then
studied in some more detail in Sec.~\ref{sec:wcrt-algebra}. For
compactness we keep these theoretical
Sections~\ref{sec:execution-schedules} and~\ref{sec:wcrt-algebra}
fairly condensed, postponing examples to Secs.~\ref{sec:examples-1}
and~\ref{sec:examples-2}. In the former, Sec.~\ref{sec:examples-1}, we
sketch applications to network flow, shortest path and task
scheduling, while in Sec.~\ref{sec:examples-2} we discuss the problem
of WCRT analysis for Esterel-style synchronous processing. The paper
concludes in Sec.~\ref{sec:conclusion} with a discussion of related
work.

\section{Syntax and Semantics of Synchronous Scheduling Interfaces} 
\label{sec:execution-schedules}

Synchronous scheduling assumes that all dependencies in the control
flow of a single instant are acyclic and the propagation of control,
for all threads, is a monotonic process in which each atomic control
point is only ever activated at most once.  Let $\vars$ be a set of
\textit{signals}, or \textit{control variables}, which specify the
atomic control points in the interface of a synchronous module.  An
\emph{event} is a subset $E \subseteq \vars$ of control variables.
A \emph{synchronous activation sequence}, or simply an
\emph{activation}, is a monotonically increasing function $\sigma \in
\nset{n} \to 2^\vars$ from $\nset n = \{ 0, 1, \ldots, n-1 \}$ into
the set of events, i.e., $\sigma(i) \subseteq \sigma(j)$ for all $0
\leq i \leq j < n$. The \emph{length} $\length{\sigma}$ of $\sigma$ is
the number of events it contains, i.e., $\length{\sigma} = \nset n$.
The unique activation of length $\nset 0 = \emptyset$ is called the
\emph{empty} activation, also denoted $\emptyset$.

Activations model the monotonic process of signal propagation during
one synchronous instant, i.e., between two ticks of the logical clock.
They induce a Boolean valuation on the control variables in the sense
that $A \in \vars$ may be considered ``present'' for the instant if $A
\in \sigma(i)$ for some $0 \leq i < \length{\sigma}$ and ``absent''
otherwise. In the former case, index $i$ is the \textit{activation
  level} for the presence of control $A$. In general, the domain
$\nset n$ over which an activation is defined acts as a discrete
domain of quantifiable resources which are consumed by control
variables becoming active at different resource levels. 
In this way, activation
sequences give an operational understanding of truth values that is
faithful to causality and resource consumption. 
A canonical interpretation is the
temporal reading: The length $\length{\sigma}$ is the duration of the
synchronous instant, i.e., the overall reaction time, and $A \in
\sigma(i)$ means that $A$ is activated, or is present from micro-step
$i$.

\begin{definition}
   Let $\sigma \in \nset n \to 2^\vars$ be an activation.
  \begin{itemize}

  \item A \emph{sub-activation} $\sigma' \subseteq \sigma$ of
    $\sigma$ is an activation $\sigma' \in \nset m \to 2^\vars$ such
    that there exists a strictly monotonic function $f \in \nset m \to
    \nset n$ with $\sigma'(i) = \sigma(f(i))$ for all $i \in \nset m$.

  \item We write $\sigma = \sigma_1 \cup \sigma_2$ to express that
    sub-activations $\sigma_1, \sigma_2 \subseteq \sigma$ form an
    \emph{activation cover} of $\sigma$, or an \emph{interleaving
      decomposition} in the sense that each event is contained in
    $\sigma_1$ or in $\sigma_2$, i.e., $\forall i \in
    \length{\sigma}.\, \exists j = 1,2.\, \exists k \in
    \length{\sigma_j}.\, i = f_j(k)$ where $f_j$ are the index
    embeddings of $\sigma_j$, $j = 1,2$.

  \item For every $i \in \Nat$ we define the \emph{shifted} activation
    $\sigma[i,:] \rcol \nset m \to 2^\vars$, where $m \df \{ j \mid 0
    \leq j + i < n \}$ and $\sigma[i,:](j) \df \sigma(j+i)$. 
\qed
\end{itemize}
\end{definition}

A shifted activation is also a sub-activation, $\sigma[i,:] \subseteq
\sigma$.  We have $\sigma[i,:] = \emptyset$ if $\sigma = \emptyset$ or
if $i \geq \length{\sigma}$.  The shift operator is monotonic wrt
sub-activations and antitonic wrt resource level, i.e., if $\sigma'
\subseteq \sigma$ and $0 \leq i \leq j$ then $\sigma'[j,:] \subseteq
\sigma[i,:]$. This depends on strict monotonicity of the
index embedding in $\sigma' \subseteq \sigma$.

In order to model non-determinism (abstracting from internal parameters or
external environment) our interfaces are interpreted over subsets
$\Sigma$ of activation sequences, called \emph{(synchronous)
  schedules}. These schedules (of a program, a module, or any other
program fragment) will be specified by a \textit{scheduling type}
$\phi$ generated by the logical operators
\begin{eqnarray*}
  \phi &\rcol\rcol=& 
    A \;\mid\; \true \;\mid\; \false \;\mid\; \phi \wedge \phi \;\mid\; 
      \neg \phi \;\mid\; \phi \imp \phi \;\mid\; 
      \phi \vee \phi \;\mid\; \phi \oplus \phi \;\mid\; 
      \phi \otimes \phi \;\mid\; \delm{\phi}
\end{eqnarray*}
generated from control variables $A \in \vars$.  We will write $\Sigma
\models \phi$ ($\sigma \models \phi$) to say that schedule $\Sigma$
(activation $\sigma$) \textit{satisfies} the type $\phi$.  The
semantics of types is formally defined below in
Def.~\ref{def:type-semantics}.  As a type specification, each control
variable $A \in \vars$ represents the guarantee that ``{\it $A$ is
  active (the signal is present, the program label has been traversed,
  the state is activated) in all activations of $\Sigma$}''.  The
constant $\true$ is satisfied by all schedules and $\false$ only by
the empty schedule or the schedule which contains only the empty
activation. The type operators $\neg$, $\imp$ are negation and
implication. The operators $\vee$ and $\oplus$ are two forms of
logical disjunction to encode internal and external non-determinism
and $\wedge$, $\otimes$ are two forms of logical conjunction related
to true concurrency and interleaving concurrency, respectively.
Finally, $\delm$ is the operator to express resource consumption.  The
usual bracketing conventions apply: The unary operators $\neg$,
$\delm$ have highest binding power, implication $\imp$ binds most
weakly and the multiplicatives $\wedge$, $\otimes$ are stronger than
the summations $\vee$, $\oplus$.  Occasionally, bi-implication $\phi
\equiv \psi$ is useful as an abbreviation for $(\phi \imp \psi) \wedge
(\psi \imp \phi)$. Also, we note that $\neg\phi$ is
equivalent to $\phi \imp \false$.

A scheduling type $\phi$ by itself only captures the functional aspect
of an interface. To get a full interface we need to enrich types by
resource information. To this end, we associate with every scheduling
type $\phi$ a set of \textit{scheduling bounds} $\sbound{\phi}$
recursively as follows:
\begin{xalignat*}{2}
  \sbound{\false} &= \nset 1 &
  \sbound{\true} &= \nset 1 \\
  \sbound{A} &= \nset 1 &
  \sbound{\neg\phi} &= \nset 1 \\
  \sbound{\phi \wedge \psi} &= \sbound{\phi} \times \sbound{\psi} &
  \sbound{\phi \vee \psi} &= \sbound{\phi} \plus \sbound{\psi} \\
  \sbound{\phi \oplus \psi} &= \sbound{\phi} \times \sbound{\psi} &
  \sbound{\phi \imp \psi} &= \sbound{\phi} \to \sbound{\psi} \\
  \sbound{\delm{\phi}} &= \cNat \times \sbound{\phi} & 
  \sbound{\phi \otimes \psi} &= \sbound{\phi} \times
  \sbound{\psi},
\end{xalignat*}
where $\nset 1 = \{ 0\}$ is a distinguished singleton set.  Elements
of the disjoint sum $\sbound{\phi} \plus \sbound{\psi}$ are presented
as pairs $(0, f)$ where $f \in \sbound{\phi}$ or $(1, g)$ where $g \in
\sbound{\psi}$. The set $\sbound{\phi} \times \sbound{\psi}$ is the
Cartesian product of the sets $\sbound{\phi}$ and $\sbound{\psi}$ and
$\sbound{\phi} \to \sbound{\psi}$ the set of total functions from
$\sbound{\phi}$ to $\sbound{\psi}$. Intuitively, an element $f \in
\sbound{\phi}$ may be seen as a form of generalised higher-order
resource matrix for schedules of shape $\phi$.

\begin{definition}
  A \emph{scheduling interface} is a pair $f \rcol \phi$ consisting
  of a scheduling type $\phi$ and a scheduling bound $f \in
  \sbound{\phi}$. An activation $\sigma$ \emph{satisfies} an
  interface $f \rcol \phi$, or \emph{satisfies} the scheduling type
  $\phi$ \emph{with} bound $f$, written $\sigma \models f \rcol \phi$,
  according to the following inductive rules:
  \begin{center}
  \begin{tabular}{rclcl}
  $\sigma$ &$\models$& $0 \rcol \false$
  &\miff& $\length{\sigma} = 0$, i.e., $\sigma = \emptyset$ \\
  $\sigma$ &$\models$& $0 \rcol \true$ 
  &\miff& always \\
  $\sigma$ &$\models$& $0 \rcol A$ 
  &\miff& $\forall 0 \leq i < \length{\sigma} \Rightarrow
                A \in \sigma(i)$ \\
  $\sigma$ &$\models$& $(f, g) \rcol \phi \wedge \psi$ 
  &\miff& $\sigma \models f \rcol \phi$ and 
  $\sigma \models g \rcol \psi$ \\
  $\sigma$ &$\models$& $(0, f) \rcol \phi \vee \psi$
  &\miff& $\sigma \models f \rcol \phi$ \\
  $\sigma$ &$\models$& $(1, g) \rcol \phi \vee \psi$
  &\miff& $\sigma \models g \rcol \psi$ \\
  $\sigma$ &$\models$& $(f, g) \rcol \phi \oplus \psi$
  &\miff& $\sigma \models f \rcol \phi$ or 
             $\sigma \models g \rcol \psi$\\
  $\sigma$ &$\models$& $f \rcol \phi \imp \psi$ 
  &\miff& $\forall \sigma' \subseteq \sigma.\;
               \forall g \in \sbound{\phi}.\;
                 (\sigma' \models g \rcol \phi \Rightarrow
                \sigma' \models f\, g \rcol \psi)$ \\
  $\sigma$ &$\models$& $(d, f) \rcol \delm\phi$ 
  &\miff& $\length{\sigma} = 0 \mbox{ or } 
              \exists i \in \Nat.\, 0 \leq i \leq d 
             \mbox{ and } \sigma[i,:] \models f \rcol \phi$ \\
  $\sigma$ &$\models$& $(f, g) \rcol \phi \otimes \psi$
  &\miff& $\exists \sigma_1,\sigma_2 \subseteq \sigma.\;
            \sigma = \sigma_1 \cup \sigma_2$ 
   $\mbox{ and }
   \sigma_1 \models f \rcol \phi \mbox{ and }
               \sigma_2 \models g \rcol \psi$. 
  \end{tabular}
  \end{center}
  A schedule $\Sigma$ \emph{satisfies} $\phi$ \emph{with} bound $f$,
  written $\Sigma \models f \rcol \phi$, if for all $\sigma \in
  \Sigma$, $\sigma \models f \rcol \phi$.  A schedule \emph{satisfies}
  $\phi$ or is \emph{bounded} for $\phi$ if there exists $f \in
  \sbound{\phi}$ such that $\Sigma \models f \rcol \phi$. \qed
\label{def:type-semantics}
\end{definition}

The semantics $\Sigma \models f \rcol \phi$ as formalised in
Def.~\ref{def:type-semantics} is a ternary relation: It links
schedules, types and bounds.  The symbol $\models$ separates the
behavioural model $\Sigma$ from the formal interface $f \rcol \phi$.
The latter, in turn, combines a qualitative and a quantitative aspect.
The type $\phi$ captures the causal relationships between the control
points and the bound $f \in \sbound{\phi}$ refines this quantitatively
by weaving in concrete activation levels.  The colon $\rcol$ is a
binary connective which separates these concerns.

\begin{proposition}
  $\sigma \models f \rcol \phi$ and $\sigma' \subseteq \sigma$ implies
  $\sigma' \models f \rcol \phi$. Moreover, $\length{\sigma} = 0$
  implies $\sigma \models f \rcol \phi$.
\label{prop:monotonicity-1}
\end{proposition}

Prop.~\ref{prop:monotonicity-1} says that interfaces are inherited by
sub-activations. This is natural since a sub-activation selects a
subset of events and thus (in general) contains more control variables
with lower activation distances. The degenerated case is the empty
activation which is inconsistent and thus satisfies all interfaces,
including the strongest specification $0 \rcol \false$, viz.\
``\textit{everything is true with zero resource consumption}''.

The most general way to use the semantic relation of
Def.~\ref{def:type-semantics} is to consider the set of (typically
abstracted) activations for a given module $P$ as a schedule
$\Sigma_P$, and then determine a suitable interface for it.  Any such
$f \rcol \phi$ with $\Sigma_P \models f \rcol \phi$ may be taken as a
valid interface specification of $P$ giving a quantified behavioural
guarantee for all activations $\sigma \in \Sigma_P$ under the given
scheduling assumptions.  Ideally, we are interested in the best
fitting or \textit{tightest} interface, if such exists.  To measure
the relative strength of an interface we employ
Def.~\ref{def:type-semantics} to associate with every pair $f \rcol
\phi$ the schedule $\sem{f \rcol \phi} = \set{\sigma}{\sigma \models f
  \rcol \phi}$ which is the semantic meaning of the interface.
Interfaces may then be compared naturally.  The smaller the set of
associated activations $\sem{f \rcol \phi}$ the tighter is the
interface $f \rcol \phi$.  Formally, we write
\begin{alignat*}{3}
   f \rcol \phi &\preceq g \rcol \psi &\qquad\mbox{if}\qquad
      \sem{f \rcol \phi} &\subseteq \sem{g \rcol \psi}
\end{alignat*}
and $f \rcol \phi \cong g \rcol \psi$ in case $\sem{f \rcol \phi} =
\sem{g \rcol \psi}$.  We call an interface $f \rcol \phi$ \emph{tight}
for $\Sigma_P$ if it is minimal wrt $\preceq$, i.e., whenever $g \rcol
\psi \preceq f \rcol \phi$ and $\Sigma_P \models g \rcol \psi$ then $f
\rcol \phi \cong g \rcol \psi$.  A tight interface provides exact
information about $\Sigma_P$ in both the functional and the resource
dimensions within the expressiveness of our typing language.
Typically, however, we are given some schedule $\Sigma_P$ together
with a \textit{fixed} type $\phi$ and ask for a minimal bound $f$ such
that $\Sigma_P \models f \rcol \phi$. If such a tight bound exists and
is unique we call it \textit{worst-case for} $\phi$.

We generalise equivalence to arbitrary types, taking $\phi \cong \psi$
to mean that for every $f \in \sbound{\phi}$ there is $g \in
\sbound{\psi}$ such that $f \rcol \phi \cong g \rcol \psi$ and vice
versa, for each $g \in \sbound{\psi}$ we can find $f \in
\sbound{\phi}$ with $g \rcol \psi \cong f \rcol \phi$.  The main
purpose of the relations $\preceq$ and $\cong$ is to justify
strengthening, weakening or semantics-preserving, transformations to
handle interfaces as tightly as sensible. They are the basis of the
interface algebra, some of whose laws will be studied next.

\section{The Algebra of Scheduling Types}
\label{sec:wcrt-algebra}

The set of scheduling bounds $\sbound{\phi}$ captures the amount of
resource information associated with a type $\phi$.  In this respect
the most simple class of types is that for which $\sbound{\phi}$ is
(order) isomorphic to $\nset 1$.  Such types are called \textit{pure}
since they do not carry resource information and thus specify only
functional behaviour.  It will be convenient to exploit the
isomorphisms $\sbound{\zeta} \cong \nset 1$ and identify all bounds $f
\in \sbound{\zeta}$ of a pure type canonically with the unique $0 \in
\nset 1$. Further, since it is unique, we may as well drop the
(non-informative) bound and simply write $\zeta$ instead of $0 \rcol
\zeta$. This means, e.g., that $\zeta_1 \wedge \zeta_2$, $(0, 0) \rcol
\zeta_1 \wedge \zeta_2$ and $0 \rcol \zeta_1 \wedge \zeta_2$ are all
identified.

Second, with this simplification on pure types in place, we may mix
bounds and types and apply the type operators to full interfaces.
Since $f \rcol \phi$ specifies individual activations it formally
behaves like an atomic statement. Hence, it is possible to use
interfaces $f \rcol \phi$ themselves as generalised ``control
variables'' in types such as $(f \rcol \phi) \wedge \psi$ or $\delm (f
\rcol \phi)$. We simply define
\begin{alignat*}{3}
 &\sbound{f \rcol \phi} \df \nset 1  &\qquad
 & \sigma \models 0 \rcol (f \rcol \phi) 
    \mbox{ iff } \sigma \models f \rcol \phi
\end{alignat*}
which turns an interface $f \rcol \phi$ into a pure type.  Then, e.g.,
$\sem{f \rcol \phi \wedge g \rcol \psi} = \sem{(0,0) \rcol (f \rcol
  \phi \wedge g \rcol \psi)} = \sem{0 \rcol (f \rcol \phi)} \cap
\sem{0 \rcol (g \rcol \psi)} = \sem{f \rcol \phi} \cap \sem{g \rcol
  \psi}$.

A few basic facts about the interface algebra arising from
Def.~\ref{def:type-semantics} are readily derived.  Not really
surprisingly, $\true$ and $\false$ are complements, $\neg\true \cong
\false$, $\neg\false \cong \true$ as well as neutral $\false \otimes
\phi \cong \false \oplus \phi \cong \true \wedge \phi \cong \phi$ and
dominant elements $\false \wedge \phi \cong \false$, $\true \oplus
\phi \cong \true \vee \phi \cong \true \otimes \phi \cong \true$.
Shifting a type by $-\infty$ and $+\infty$ produces the strongest and
weakest statements $\false$ and $\true$, respectively:

\begin{proposition}
  For arbitrary types $\phi$, $-\infty \rcol \delm \phi \cong \false$
  and $+\infty \rcol \delm\phi \cong \true$. 
   \label{prop:infinite-shift}
\end{proposition}

All operators $\vee$, $\wedge$, $\oplus$ and $\otimes$ are
commutative. The pairs $\vee \leftrightarrow \wedge$ and $\oplus
\leftrightarrow \wedge$ fully distribute over each other, while
$\otimes$ distributes over both $\oplus$ and $\vee$, but not the other
way round. Between $\otimes$ and $\wedge$ no distribution is possible,
in general. One can show that the fragment $\vee$, $\wedge$,
$\false$, $\neg$, $\imp$ satisfies the laws of Heyting algebras seen
in Prop.~\ref{prop:heyting-laws}.

\begin{proposition}
  For arbitrary types $\phi_1$, $\phi_2$, $\psi$:
  \begin{xalignat*}{2}
     \psi \imp \psi &\cong \true &\quad
     \phi_1 \imp (\phi_2 \imp \phi_1) &\cong \true \\
     (\phi_1 \wedge \phi_2) \imp \psi &\cong 
       \phi_1 \imp (\phi_2 \imp \psi) & \quad 
     (\phi_1 \imp \phi_2) \wedge \phi_1 &\cong 
       \phi_1 \wedge \phi_2 \\
       (\phi_1 \vee \phi_2) \imp \psi &\cong 
       (\phi_1 \imp \psi) \wedge (\phi_2 \imp \psi) &\quad
      \psi \imp (\phi_1 \wedge \phi_2) &\cong
         (\psi \imp \phi_1) \wedge (\psi \imp \phi_2) \\
      \false \imp \psi &\cong \true &\quad
      \psi \imp \true &\cong \true \\
      \psi \imp \false &\cong \neg\psi &\quad
      \true \imp \psi &\cong \psi.
  \end{xalignat*} \qed
   \label{prop:heyting-laws}
\end{proposition}

\vspace{-0.6cm}

It is worthwhile to observe that the classical principles of the
Excluded Middle $A \oplus \neg A$ and $A \vee \neg A$ are both
different and not universally valid in WCRT algebra. The latter says
$A$ is \textit{static}, i.e., $A$ is present in all activations or
absent in all activations, the former that signal $A$ is
\textit{stable}, i.e., in each activation individually, $A$ is either
present from the start or never becomes active. Clearly, not every
signal is static or stable.  The absence of the axioms $A \oplus \neg
A$, $A \vee \neg A$, which arises naturally from the activation
semantics, is a definitive characteristics of intuitionistic logic or
Heyting algebra.  This feature is crucial to handle the semantics of
synchronous languages in a compositional and fully abstract
way~\cite{mvm01-tocl}.

\paragraph{Boolean Types.}

An important sub--class of pure types are negated types $\neg\phi$.
They express universal statements about each singleton event of each
activation sequence in a schedule. For instance, $\Sigma \models
\neg(A \otimes B)$ says that no event $\sigma(i) \subseteq \vars$ ($0
\leq i < \length{\sigma}$) in any $\sigma \in \Sigma$ contains $A$ or
$B$. Similarly, $\neg(A \imp B)$ states that $A$ is present and $B$ is
absent in every event of every activation sequence, which is the same
as $\neg\neg(A \wedge \neg B)$.  Negated types are expressively
equivalent to, and can be transformed into, \textit{Boolean} types
obtained from the following grammar, where $\phi$ is an arbitrary
type:
\begin{eqnarray*}
  \beta &\rcol\rcol=&
  \true \;\mid\; \false \;\mid\; A \;\mid\; \neg\beta \;\mid\; 
   \beta \wedge \beta \;\mid\; \beta \otimes \beta \;\mid\;
  \phi \imp \beta.
\end{eqnarray*}

\begin{proposition}
  The Boolean types form a Boolean algebra with $\neg$, $\wedge$,
  $\otimes$ as classical complement, conjunction and disjunction,
  respectively. Moreover, $\Sigma \models \beta$ iff for every $\sigma
  \in \Sigma$ and $i \in \length{\sigma}$ the event $\sigma(i)
  \subseteq \vars$ satisfies $\beta$ as a classical Boolean formula in
  control variables $\vars$. \qed
 \label{prop:boolean-types}
\end{proposition}

A consequence of Prop.~\ref{prop:boolean-types} is that the interface
algebra contains ordinary classical Boolean algebra as the fragment of
Boolean types. In particular, for Boolean types the Double Negation
principle $\neg\neg\beta \cong \beta$ and Excluded Middle $\neg\beta
\otimes \beta \cong \true$ hold as well as the De-Morgan Laws
$\neg(\beta_1 \wedge \beta_2) \cong \neg\beta_1 \otimes \neg\beta_2$
and $\neg(\beta_1 \otimes \beta_2) \cong \neg\beta_1 \wedge
\neg\beta_2$.  Boolean types, like all types satisfying $\neg\neg\phi
\cong \phi$ or $\neg\phi \otimes \phi \cong \true$, behave exactly
like expressions of Boolean algebra, encapsulating a Boolean condition
to be satisfied by each event in a sequence.

\paragraph{Pure Types.}

The sum operator $\oplus$ takes us outside the sub-language of Boolean
types.  The reason is that the truth of $\oplus$, e.g., in stability
$A \oplus \neg A$, depends on the global behaviour of an activation
and cannot be reduced to a \textit{single} Boolean condition.  This is
highlighted by the difference between $\sigma \models A \oplus B$
which is the condition $\forall i \in \length{\sigma},\, A \in
\sigma(i) \mbox{ or } \forall i \in \length{\sigma},\, B \in
\sigma(i)$ and $\sigma \models A \otimes B$ which says $\forall i \in
\length{\sigma},\, A \in \sigma(i) \mbox{ or } B \in \sigma(i)$. The
larger class of pure types, which includes $\oplus$, give us the
possibility to express ``Boolean'' conditions \textit{across}
activations, as opposed to Boolean types which act \textit{within}
activations. The pure types, denoted by meta-variable $\zeta$, are
characterised syntactically as follows:
\begin{eqnarray*}
  \zeta &\rcol\rcol=&
  \beta \;\mid\; 
  \zeta \wedge \zeta \;\mid\; 
  \zeta \oplus \zeta \;\mid\; \zeta \otimes \zeta \;\mid\; \phi \imp \zeta, 
\end{eqnarray*}
where $\beta$ is Boolean and $\phi$ is an arbitrary type. Notice that
not only every Boolean type, but also every negation $\neg\phi = \phi
\imp \false$, is pure according to this syntactic criterion.

\begin{proposition}
  Every pure type $\zeta$ has a representation $\zeta \cong
  \bigoplus_i \beta_{i}$ over Boolean types $\beta_{i}$.
\label{prop:pure-normalise}
\end{proposition}

\paragraph{Elementary Types.}

Pure types have the special property that schedules $\Sigma$ are
bounded for them iff each individual activation $\sigma \in \Sigma$
is bounded, i.e., they express properties of individual
activations.  Formally, if $\Sigma_1 \models \zeta$ and $\Sigma_2
\models \zeta$ then $\Sigma_1 \cup \Sigma_2 \models \zeta$.
Disjunctions $\zeta_1 \vee \zeta_2$ and resource types $\delm\zeta$,
in contrast, do not share this locality property: Although each
activation $\sigma$ may satisfy $\zeta_1$ or $\zeta_2$, the schedule
$\Sigma$ as a whole need not be resource-bounded for $\zeta_1 \vee
\zeta_2$ as this would mean all activations satisfy $\zeta_1$ or all
satisfy $\zeta_2$. Similarly, each individual activation $\sigma \in
\Sigma$ may validate $\zeta$ with some resource bound, 
without necessarily there
being a \textit{single} common bound for all activations in $\Sigma$.

A useful class of types containing $\vee$ and $\delm$ are those for
which $\sbound{\phi}$ is canonically or\-der-isomor\-phic to a Cartesian
product of numbers, i.e., to $\cNat^{\nset n}$ for some $n \geq 0$.
These scheduling types $\phi$ with $\sbound{\phi} \cong \cNat^{\nset
  n}$ are called \textit{elementary}.  They are generated by the
grammar
\begin{eqnarray*}
  \theta &\rcol\rcol=& 
      \zeta \;\mid\; 
      \theta \wedge \theta \;\mid\; \theta \oplus \theta \;\mid\; 
       \theta \otimes \theta \;\mid\;
      \delm{\zeta} \;\mid\; \psi \imp \theta,
\end{eqnarray*}
where $\zeta$ is pure and $\psi$ is $\delm$-free. Elementary
scheduling types are of special interest since their elements are
first-order objects, i.e., vectors and matrices of natural numbers.

Elementary interfaces specify the resource consumption of logical
controls. For instance, $\sigma \models (d,0) \rcol \delm\zeta$, given
$\zeta = \oplus_i \beta_i$ (see Prop.~\ref{prop:pure-normalise}), says
that $\sigma$ enters and remains inside a region of events 
described by one of the
Boolean conditions $\beta_i$ and consumes at most $d$ resource units
to do that. The special case $\sigma \models d \rcol \delm\false$ says
that $\sigma$ consumes no more than $d$ units during any instant.
Similarly, $\sigma \models \zeta \imp (d,0) \rcol \delm \xi$ with
$\zeta = \oplus_i \beta_i$ and $\xi = \oplus_j \gamma_j$ says that
every sub-activation $\sigma' \subseteq \sigma$ that runs
fully inside one of the regions $\beta_i$ must reach one of the
regions $\gamma_j$ with resources bounded by $d$.  Then, $\sigma
\models \zeta \imp (d,0) \rcol \delm \false$ means that $\sigma$
consumes no more than $d$ units while staying in any of the regions $\beta_i$.

To compactify the notation we will write tuples $(d_1, d_2)$ for the
bounds $((d_1,0), (d_2,0)) \in (\cNat \times \nset 1) \times (\cNat
\times \nset 1) \cong \cNat \times \cNat$ of types such as $\delm
\zeta_1 \oplus \delm \zeta_2$, $\delm \zeta_1 \wedge \delm \zeta_2$,
$\delm \zeta_1 \otimes \delm \zeta_2$.  We apply this
simplification also to bounds $f \in \nset 1 \to \cNat \times \nset 1
\cong \cNat$ for types such as $\zeta_1 \imp \delm \zeta_2$: We write
$[d] \rcol \zeta_1 \imp \delm \zeta_2$, treating the bracketed value
$[d]$ like a function $\lambda x.\, (d,0)$. In fact, $[d] \rcol
\zeta_1 \imp \delm\zeta_2$ is the special case of a $1 \times 1$
matrix.  We will systematically write \textit{column vectors} $[d_1;
d_2]$ instead of $\lambda x. ((d_1,0), (d_2,0))$ for the bounds of
types such as $\zeta \imp \delm\zeta_1 \oplus \delm\zeta_2$, $\zeta
\imp \delm\zeta_1 \wedge \delm\zeta_2$ or $\zeta \imp \delm\zeta_1
\otimes \delm\zeta_2$, and \textit{row-vectors} $[d_1, d_2]$ in place
of $\lambda x.\, \mbox{case}\, x\, \mbox{of}\, [(0,0) \to (d_1,0),
(1,0) \to (d_2,0)]$ for types $\zeta_1 \vee \zeta_2 \imp
\delm\zeta$. Our linearised matrix notation uses semicolon for
row-wise and ordinary colon for columns-wise composition of
sub-matrices.  Specifically, $[d_{11}; d_{21}, d_{12}; d_{22}]$ and
$[d_{11}, d_{12}; d_{21}, d_{22}]$ denote the same $2 \times 2$
matrix.

In the following Secs.~\ref{sec:examples-1} and~\ref{sec:examples-2}
we are going illustrate different sub-algebras of specialised
elementary types to manipulate combined functional and quantitative
information and to facilitate interface abstractions. These generalise
the algebra of dioids~\cite{BaccelliCOQ:SyncNLin,LeeZZ05} to full
max-min-plus, obtaining an equally tight as uniform combination of
scheduling algebra and logical reasoning.

\section{Examples I: Network Flow, Shortest Path, Task Scheduling}
\label{sec:examples-1}

The logical operations on types control the arithmetical operations on
resource bounds.  The next two
Props.~\ref{prop:basic-inequations} and~\ref{prop:tensor} sum up some
important basic facts.

\begin{proposition}
  The arithmetic operations $\mind$, $\maxd$ and $+$ compute
  worst-case bounds such that
  \begin{alignat}{3}
    &[d_1] \rcol \zeta_1 \imp \delm \zeta_2 
      &\; &\wedge [d_2] \rcol \zeta_2 \imp \delm \zeta_3 &\; &\;\preceq\; 
      [d_1 + d_2] \rcol \zeta_1 \imp  \delm\zeta_3
   \label{eqn:sequential-comp} \\
    &[d_1] \rcol \zeta \imp \delm\zeta_1  
        &\; &\wedge [d_2] \rcol \zeta \imp \delm\zeta_2 &\; &\;\preceq\;
      [\maxd(d_1,d_2)] \rcol \zeta \imp \delm(\zeta_1 \wedge \zeta_2) 
   \label{eqn:and-abstr-out} \\
    &[d_1] \rcol \zeta \imp \delm\zeta_1  
        &\; &\wedge [d_2] \rcol \zeta \imp \delm\zeta_2 &\; &\;\preceq\;
      [\mind(d_1,d_2)] \rcol \zeta \imp \delm(\zeta_1 \oplus \zeta_2)
   \label{eqn:sum-comp-out} \\
    &[d_1] \rcol \zeta_1 \imp \delm\zeta 
        &\; &\wedge [d_2] \rcol \zeta_2 \imp \delm\zeta &\; &\;\preceq\;
      [\maxd(d_1,d_2)] \rcol (\zeta_1 \oplus \zeta_2) \imp \delm\zeta 
   \label{eqn:sum-abstr-in} \\
   &[d_1] \rcol \zeta_1 \imp \delm\zeta 
        &\; &\wedge [d_2] \rcol \zeta_2 \imp \delm\zeta &\; &\;\preceq\;
      [\mind(d_1,d_2)] \rcol (\zeta_1 \wedge \zeta_2) \imp \delm\zeta.
   \label{eqn:and-comp-in}
  \end{alignat}
  \label{prop:basic-inequations}
\end{proposition}

\vspace{-0.6cm}

The law \eqref{eqn:sequential-comp} expresses a \textit{sequential
  composition} of an offset by $d_1$ from control point $\zeta_1$ to
$\zeta_2$ with a further shift of $d_2$ from $\zeta_2$ to $\zeta_3$.
The best guarantee we can give for the cost between $\zeta_1$ and
$\zeta_3$ is the addition $d_1 + d_2$. The bounds $[d_1]$
and $[d_2]$ act like typed functions with $[d_1 + d_2]$ being function
composition, $[d_2] \cdot [d_1] = [d_1 + d_2]$. This is nothing but
the multiplication of $1 \times 1$ matrices in max-plus or min-plus
algebra.  The law \eqref{eqn:and-abstr-out} is \textit{conjunctive
  forking}: If it takes at most $d_1$ units from $\zeta$ to some
control point $\zeta_1$ \textit{and} at most $d_2$ to $\zeta_2$, then
we know that within $\maxd(d_1, d_2)$ we have activated
both together, $\zeta_1 \wedge \zeta_2$. A special case of this occurs
when $\zeta \cong \true$, i.e., $d_1 \rcol \delm\zeta_1 \wedge d_2
\rcol \delm\zeta_2 \cong \maxd(d_1, d_2) \rcol \delm(\zeta_1 \wedge
\zeta_2)$. Now suppose conjunction is replaced by sum $\zeta_1 \oplus
\zeta_2$, i.e., we are only interested in activating one of $\zeta_1$
\textit{or} $\zeta_2$, but do not care which.  The worst-case bound
for this \textit{disjunctive forking} is the minimum, as seen in
\eqref{eqn:sum-comp-out}. Again, there is the special case $d_1 \rcol
\delm\zeta_1 \wedge d_2 \rcol \delm\zeta_2 \cong \mind(d_1, d_2) \rcol
\delm(\zeta_1 \oplus \zeta_2)$.  Dually, \textit{disjunctive joins}
\eqref{eqn:sum-abstr-in} are governed by the maximum: Suppose that
starting in $\zeta_1$ activates $\zeta$ with at most $d_1$ cost and
starting in $\zeta_2$ takes at most $d_2$ resource units.  Then, if we
only know the activation starts from $\zeta_1$ \textit{or} $\zeta_2$
but not which, we can obtain $\zeta$ if we are prepared to expend the
maximum of both costs. If, however, we assume the schedule activates
both $\zeta_1$ \textit{and} $\zeta_2$, which amounts to
\textit{conjunctive join}, then the destination $\zeta$ is obtained
with the minimum of both shifts, see \eqref{eqn:and-comp-in}.

\begin{proposition}
  Let $\zeta_1$, $\zeta_2$ be pure types which are persistent in the
  sense that whenever $\sigma(k) \models \zeta_i$ for $0 \leq k <
  \length{\sigma}$, then $\sigma[k,:] \models \zeta_i$, too. Then,
  \begin{xalignat}{1}
     & d_1 \rcol \delm\zeta_1  \otimes d_2 \rcol \delm\zeta_2
         \;\preceq\;  d_1 + d_2 \rcol \delm(\zeta_1 \oplus \zeta_2) 
    \label{eqn:simple-otimes} \\
         & (d_1 \rcol \delm\zeta_1  \wedge (\zeta_1 \imp \zeta_2))
         \otimes (d_2 \rcol \delm\zeta_2 \wedge (\zeta_2 \imp \zeta_1))
     \;\preceq\;  d_1 + d_2 \rcol \delm(\zeta_1 \wedge \zeta_2). 
   \label{eqn:sync-otimes}
 \end{xalignat}
\label{prop:tensor}
\end{proposition}

\vspace{-0.6cm}

Consider~\eqref{eqn:simple-otimes} of Prop.~\ref{prop:tensor}.
Suppose a schedule $\sigma$ splits into two (sub-)threads $\sigma =
\sigma_1 \cup \sigma_2$ each switching control $\zeta_1$ and $\zeta_2$
consuming at most $d_1$ and $d_2$ units, respectively.  Since they can
be arbitrarily interleaved and we do not know which one completes
first, all we can claim is $\sigma(k) \models \zeta_i$ for some $k
\leq d_1 + d_2$ and $i = 1,2$. By persistence, this suffices to
maintain $\zeta_i$ from level $k$ onwards, so that $\sigma \models d_1
+ d_2 \rcol \delm(\zeta_1 \oplus \zeta_2)$.
Without imposing further assumptions, a sub-thread may be allocated an
unknown number of resource units, thereby stalling the progress of the
other, unboundedly.  The situation changes, however, if the $\zeta_i$
are \textit{synchronisation points} where the threads must
give up control unless the other thread has passed its own
synchronisation point $\zeta_j$ ($i \neq j$), too.  This is the content
of~\eqref{eqn:sync-otimes} and specified formally by the additional
constraints $\zeta_i \imp \zeta_j$.

Prop.~\ref{prop:basic-inequations} and~\ref{prop:tensor} highlight how
the arithmetic of min-max-plus algebra are guided by the logical semantics
of interface types. From this vantage point, resource analysis
is nothing but a semantics-consistent manipulation of a collection of
numbers: Whether $[d_1] \rcol \phi_1$, $[d_2] \rcol \phi_2$ are to be
added, maximised or minimised depends on their types $\phi_1$ and
$\phi_2$. In particular, keeping track of the types will make the
difference between a max-of-sums (sum-of-mins) as opposed to a
sum-of-maxes (min-of-sums).

\subsection{Network Flow}
\label{ex:max-flow-prob}

\begin{wrapfigure}[10]{r}{8cm}
  \centering
\includegraphics{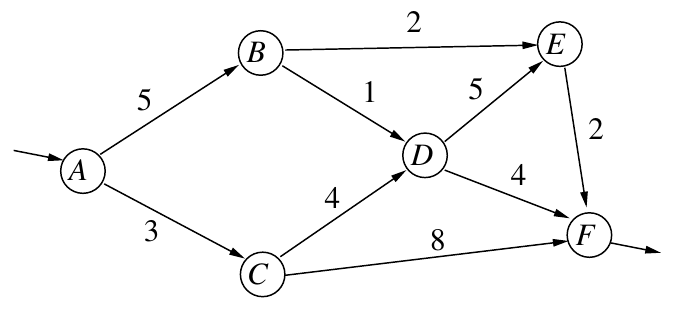}%
  \caption{Scheduling Dependency Graph $N$}
  \label{fig:sched-dep-grf}
\end{wrapfigure}

Consider the dependency graph in Fig.~\ref{fig:sched-dep-grf} with
control nodes $\vars = \{A, B, C, D, E, F\}$ and dependency edges
labelled by positive integers.  Let us assume the graph models a
communication network in which control nodes represent packet routers
and edges are directed point-to-point connections of limited
bandwidth. For instance, the router at node $D$ receives packets from
routers $B$ and $C$ through channels of bandwidth $1$ and $4$,
respectively. It forwards the incoming traffic to routers $E$ or $F$
of bandwidth $5$ and $4$, respectively. The bandwidth measures the
maximal amount of information that can travel across the channel per
synchronisation instant.  The analysis of the maximum throughput is a
synchronous scheduling problem which can be modelled using interface
types.

We associate with the network $N$ a scheduling type $\phi_N$, such
that the amount of packets that can be pushed into a node $X$ is given
by the minimal $d$ such that $\phi_N \preceq X \imp d \rcol
\delm\false$, i.e., the maximal number of scheduling cycles that node
$X$ may be kept alive within any activation specified by $\phi_N$. The
idea is that if $\sigma \in \sem{\phi_N}$ is a valid activation of $N$
then each cycle $i \in \length{\sigma}$ such that $X \in \sigma(i)$
represents a packet unit $i$ sent through $X$. The event $\sigma(i)
\subseteq \vars$ encodes the packet's path, i.e., the set of all
routers that payload unit $i$ is passing on its journey through the
network.  The statement $\sigma \models X \imp d \rcol \delm \false$
then says that whenever $X$ becomes alive in activation $\sigma$ it
handles no more $d$ packets.  This number may vary between
activations.  The minimal $d$, bounding all activations in this way,
is the maximal throughput at $X$ permitted by specification $\phi_N$.
Observe that both capacity values $0$ and $-\infty$ are equivalent, $0
\rcol \delm \false \cong -\infty \rcol \delm \false \cong \false$.  In
fact, the type $X \imp 0 \rcol \delm\false$ paraphrased ``$X$ forwards
$0$ packets'' and $X \imp -\infty \rcol \delm \false$ saying ``$X$
does not forward any packets'', are the same statements and equivalent
to $\neg X$.

Now consider node $D$ again. Within the synchronous measurement
instant, all packets arriving at $D$ must be scheduled to leave
through channels $D \to E$ or $D \to F$. Consider an activation
$\sigma \models D$, i.e., all $i \in \length{\sigma}$ are packets
dispatched through $D$. Some of these will go to $E$, others to $F$
and all go to one of the two. Hence there are sub-activations
$\sigma = \sigma_1 \cup \sigma_2$ such that $\sigma_1 \models E$
and $\sigma_2 \models F$. Also, because of the channel limitations,
there can be at most $5$ packet units of the former and $4$ of the
latter type. Thus, $\sigma_1 \models E \wedge 5 \rcol \delm\false$ and
$\sigma_2 \models F \wedge 4 \rcol \delm\false$.  All in all, we have
found the type specifying $D$ and its connections in $N$ to be $D \imp
(E \wedge 5 \rcol \delm\false) \otimes (F \wedge 4 \rcol
\delm\false)$.

The tensor $\otimes$ is used to model the output branching at a node.
Observe that if we increase one of the channel capacities to
$+\infty$, say the one giving access to $E$, we get $D \imp (E \wedge
+\infty \rcol \delm\false) \otimes (F \wedge 4 \rcol \delm\false)
\cong D \imp E \otimes (F \wedge 4 \rcol \delm\false)$ because $E
\wedge +\infty \rcol \delm\false \cong E \wedge \true \cong E$. This
means the channel $D \to E$ does not impose any further constraints on
the throughput besides what $E$ prescribes.  If we decrease the
capacity to $0$, the type reduces to $D \imp (E \wedge 0 \rcol
\delm\false) \otimes (F \wedge 4 \rcol \delm\false) \cong D \imp F
\wedge 4 \rcol \delm\false$ since $E \wedge 0 \rcol \delm\false \cong
E \wedge \false \cong \false$ and $\false \otimes \phi \cong \phi$.
Hence, a capacity of $0$ behaves as if the channel was cut off
completely.  Consequently, the degenerated case of a node $X$ without
any exits would be specified by $X \imp \false$ or $\neg X$.
If we conjoin the types for all nodes of $N$ as seen in
Fig.~\ref{fig:sched-dep-grf}, we get
  \begin{alignat}{1}
    \phi_N \df 
       &\quad\phantom{\wedge}  
           \true \imp (A \wedge +\infty \rcol \delm\false) 
     \label{eqn:flow-src-eq} \\
     &\quad\wedge 
        A \imp ((B \wedge 5 \rcol \delm\false) \otimes 
                     (C \wedge 3 \rcol \delm\false)) 
     \label{eqn:flow-A-eq} \\
     &\quad\wedge 
         B \imp ((E \wedge 2 \rcol \delm\false) \otimes 
             (D \wedge 1 \rcol \delm\false)) 
     \label{eqn:flow-B-eq} \\
     &\quad\wedge
         C \imp ((D \wedge 4 \rcol \delm\false) \otimes 
             (F \wedge 8 \rcol \delm\false)) 
     \label{eqn:flow-C-eq} \\
     &\quad\wedge  
         D \imp ((E \wedge 5 \rcol \delm\false) \otimes 
             (F \wedge 4 \rcol \delm\false)) 
     \label{eqn:flow-D-eq} \\
     &\quad\wedge 
         E \imp (F \wedge 2 \rcol \delm\false) 
     \label{eqn:flow-E-eq} \\
     &\quad\wedge
         F \imp (\true \wedge +\infty \rcol \delm\false).
     \label{eqn:flow-F-eq} 
  \end{alignat}
  Type~\eqref{eqn:flow-src-eq} designates $A$ as the source node
  of the network. It formalises a source channel of infinite capacity
  permitting the global environment, represented by the logical
  control $\true$, to push as many packets as possible into $A$.
  Analogously, destination node $F$~\eqref{eqn:flow-F-eq} returns
  packets back to the external environment. Again, this sink channel
  has infinite capacity, since all packets arriving at
  $F$ will delivered.

  The throughput $d_N$ of $N$ is the smallest $d$ such that $\phi_N
  \preceq d \rcol \delm\false$.  To get the ``exact'' or ``optimal''
  bound we must explore the network in breadth and depth.  The
  analysis strategy involves non-linear global optimisation such as
  the Ford-Fulkerson or Goldberg's Preflow-Push algorithms.  This is
  not the place to review these algorithm. We shall merely indicate
  how their logical content can be coded in type theory.  Consider
  that each of the network implications
  ~\eqref{eqn:flow-src-eq}--\eqref{eqn:flow-F-eq} of the form $X \imp
  \otimes_Y (Y \wedge d_Y \rcol \delm\false)$ can be used as an
  equation $X \cong X \wedge \otimes_Y (Y \wedge d_Y \rcol
  \delm\false)$ for transformations by substitution. For example,
  proceeding forwards from the source $A$, breadth-first, we can
  derive
  \begin{alignat}{1}
    \true &\cong A 
      \nonumber \\
     &\cong\; 
         A \wedge ((B \wedge 5 \rcol \delm\false) \otimes 
                (C \wedge 3 \rcol \delm\false))
    \nonumber \\
    &\cong
       A \wedge 
          ((B \wedge ((E \wedge 2 \rcol \delm\false) \otimes 
             (D \wedge 1 \rcol \delm\false))  
                \wedge 5 \rcol \delm\false) 
    \nonumber \\
    &\qquad\qquad \otimes 
        (C \wedge ((D \wedge 4 \rcol \delm\false) \otimes 
        (F \wedge 8 \rcol \delm\false)) \wedge 3 \rcol \delm\false)) 
    \nonumber \\
    &\cong
       ((A \wedge B \wedge E \wedge 2 \rcol \delm\false)
     \label{eqn:flow-exp-eq-1} \\
    &\qquad  \otimes (A \wedge B \wedge D \wedge 1 \rcol \delm\false)) 
     \label{eqn:flow-exp-eq-2} \\
    &\qquad\quad \otimes 
        (((A \wedge C \wedge D \wedge 3 \rcol \delm\false) 
    \label{eqn:flow-exp-eq-3} \\
    &\qquad\qquad \otimes 
        (A \wedge C \wedge F \wedge 3 \rcol \delm\false)) \wedge 
            3 \rcol \delm\false),
    \label{eqn:flow-exp-eq-4}
  \end{alignat}
  using the special $\wedge/\otimes$ distribution $X \wedge (\phi_1
  \otimes \phi_2) \cong (X \wedge \phi_1) \otimes (X \wedge \phi_2)$
  for atoms $X \in \vars$, and the derivable laws $((\phi_1 \wedge d_1
  \rcol \delm\false) \otimes (\phi_2 \wedge d_2 \rcol \delm\false))
  \wedge e \rcol\delm\false \cong (\phi_1 \wedge d_1 \rcol
  \delm\false) \otimes (\phi_2 \wedge d_2 \rcol \delm\false)$ for $e
  \geq d_1 + d_2$ and $((\phi_1 \wedge d_1 \rcol \delm\false) \otimes
  (\phi_2 \wedge d_2 \rcol \delm\false)) \wedge e \rcol\delm\false
  \cong (\phi_1 \wedge e \rcol \delm\false) \otimes (\phi_2 \wedge e
  \rcol \delm\false) \wedge e \rcol \delm\false$ for $e \leq
  \mind(d_1,d_2)$.
  
  The type~\eqref{eqn:flow-exp-eq-1}--\eqref{eqn:flow-exp-eq-4}
  describes the resource usage of packets entering the network up to a
  depth of $3$ nodes, classifying them into 4 separate flows: The
  packets from~\eqref{eqn:flow-exp-eq-1} pass through $A \to B \to E$
  and can occupy at most $2$ bandwidth units, those
  from~\eqref{eqn:flow-exp-eq-2} follow the path $A \to B \to D$ and
  have a volume of at most $1$ unit.  Furthermore, the
  packets~\eqref{eqn:flow-exp-eq-3} travelling along $A \to C \to D$
  or~\eqref{eqn:flow-exp-eq-4} on path $A \to C \to F$ each have at
  most volume $3$, as specified by $A \wedge C \wedge D \wedge 3 \rcol
  \delm\false$ and $A \wedge C \wedge F \wedge 3 \rcol \delm\false$.
  Moreover, their sum must not exceed the limit $3$ either, as
  enforced by the extra outer conjunct $3 \rcol \delm\false$.  The
  maximal flow through the network can be obtained by applying the
  (in-)equations~\eqref{eqn:flow-exp-eq-1}--\eqref{eqn:flow-exp-eq-4}
  in this fashion until saturation is achieved, when all logical
  controls may be dropped, turning equation $\cong$ into inequation
  $\preceq$:
  \begin{alignat*}{1}
    \true &\cong\; A \;\cong\; \cdots \\
    &\cong
       ((A \wedge B \wedge E \wedge F \wedge 2 \rcol \delm\false) \\
    &\qquad \otimes 
        (A \wedge B \wedge D \wedge F \wedge 1 \rcol \delm\false)) \\
    &\qquad\quad\otimes 
        (((((A \wedge C \wedge D \wedge F \wedge 3 \rcol \delm\false) \\
    &\qquad\qquad\qquad  \otimes (A \wedge C \wedge D \wedge E \wedge F 
              \wedge 2 \rcol \delm\false))
         \wedge 3 \rcol \delm\false) \\
    &\qquad\qquad\quad \otimes 
        (A \wedge C \wedge F \wedge 3 \rcol \delm\false)) \wedge 
            3 \rcol \delm\false) \\
    &\preceq
       (2 \rcol \delm\false \otimes 1 \rcol \delm\false) \\
    &\qquad\otimes 
        ((((3 \rcol \delm\false  \otimes 2 \rcol \delm\false)
         \wedge 3 \rcol \delm\false) \otimes 
        3 \rcol \delm\false) \wedge 3 \rcol \delm\false)
    \;\cong\; 6 \rcol \delm\false,
  \end{alignat*}
  using the laws $d \rcol \delm\false \wedge e \rcol \delm\false \cong
  \mind(d,e) \rcol \delm\false$ and $d \rcol \delm\false \otimes e
  \rcol \delm\false \cong d + e \rcol \delm\false$, derived from
  \eqref{eqn:sum-comp-out} and \eqref{eqn:simple-otimes}, respectively.

  This saturation process is a fixed-point construction which may be
  implemented using a standard ``max-flow'' algorithm. Specifically,
  the graph algorithms of Ford-Fulkerson or Goldberg are efficient
  decision procedures for deciding the algebra induced by the fragment
  of types appearing
  in~\eqref{eqn:flow-src-eq}--\eqref{eqn:flow-exp-eq-4}. This
  sub-algebra of ``logical numbers'' provides a purely algebraic
  interpretation for these standard algorithms. It should be clear
  that the graph-theoretic information is coded in the syntactic
  structure of the types.  However, in contrast to plain graphs, types
  are equipped with behavioural meaning in the form of scheduling
  sequences. They generate a plus-min algebra of scheduling sequences
  which is not a linear algebra, as it does not satisfy distribution.
  Specifically, $e \rcol \delm \false \wedge (d_1 \rcol \delm\false
  \otimes d_2 \rcol \delm\false) \cong \mind(e, d_1 + d_2) \rcol
  \delm\false \preceq \mind(e, d_1) + \mind(e, d_2) \rcol \delm\false
  \cong (e \rcol \delm\false \wedge d_1 \rcol\delm\false) \otimes (e
  \rcol\delm \false \wedge d_2 \rcol\delm\false)$. This approximation
  offset, of course, is why max-flow problems are not linear matrix
  problems but require global search and relaxation methods. \qed

\subsection{Shortest Path}
\label{ex:short-path-prob}

A different interpretation of the scheduling graph
Fig.~\ref{fig:sched-dep-grf} reads the edge labels as \emph{distances}
and asks for the length of the shortest path through the network. This
leads to an ``inverted'' network algebra: The sequential composition
of edges is \textit{addition} and the branching of edges at a node is
associated with the \textit{minimum} operation, whereas in the network
flow situation of Sec.~\ref{ex:max-flow-prob}, sequential composition
corresponds to minimum and branching is addition. Not surprisingly,
the shortest path interpretation invokes a different fragment of the
type theory.  Again, each node is a control variable $\vars = \{ A, B,
C, D, E, F\}$. An activation $\sigma$ models a journey through the
network activating control nodes as it passes them.  If $\sigma$
activates $X$ at time $i$, then $X \in \sigma(i)$, and if it traverses
an edge $X \to Y$ with distance label $d$, then for some $0 \leq k
\leq d$, $Y \in \sigma(i+k)$. Hence $\sigma$ satisfies the type $X
\imp d \rcol \delm Y$. If there are several outgoing edges $X \to Y_1$
and $X \to Y_2$ and $\sigma$ reaches $X$, then, because we are
interested in the \textit{shortest} path, we permit $\sigma$ to
explore both branches ``in parallel''. Hence, $\sigma$ fulfils both
implications $X \imp d_1 \rcol \delm Y_1$ and $X \imp d_2 \rcol \delm
Y_2$.  Following this idea, the network $N$ as given in
Fig.~\ref{fig:sched-dep-grf} comes out as the type specification
\begin{alignat}{1}
    \phi_N &\df A \imp 5 \rcol \delm B 
    \;\wedge\; A \imp 3 \rcol \delm C
    \;\wedge\; B \imp 1 \rcol \delm D
    \;\wedge\; B \imp 2 \rcol \delm E \nonumber \\
    &\qquad \;\wedge\; C \imp 4 \rcol \delm D
    \;\wedge\; C \imp 8 \rcol \delm F
    \;\wedge\; D \imp 5 \rcol \delm E
    \;\wedge\; D \imp 4 \rcol \delm F
    \;\wedge\; E \imp 2 \rcol \delm F.
    \label{eqn:short-path-type}
\end{alignat}
The length of the shortest path between $X$ and $Y$ is the minimal $d$
such that $\phi_N \preceq X \imp d \rcol \delm Y$.
By~\eqref{eqn:sequential-comp}, sequentially connecting edges $X \imp
d_1 \rcol \delm Y$ and $Y \imp d_2 \rcol \delm Z$ yields $X \imp d_1 +
d_2 \rcol \delm Z$, and a choice of two paths $X \imp d_1 \rcol \delm
Z$ and $X \imp d_2 \rcol \delm Z$ between the same start and end node,
by~\eqref{eqn:sum-comp-out} implies $X \imp \mind(d_1, d_2) \rcol
\delm Z$ as desired.  Now the values of $0$ and $-\infty$ have
different meaning: $X \imp 0 \rcol \delm Y$ is equivalent to $X \imp
Y$ modelling an edge without cost. In contrast, $X \imp -\infty \rcol
\delm Y$ is semantically the same as $X \imp \false$ which says that
no activation reaches control node $X$.  A distance $+\infty$
expresses absence of a connection since $X \imp +\infty \rcol \delm Y
\cong X \imp \true \cong \true$ which does not give any information
about how to reach $Y$ from $X$.

It is well-known how to compute shortest paths by linear programming.
This exploits the distribution law $\min(e + d_1, e + d_2) = e +
\mind(d_1, d_2)$, which permits us to organise the scheduling bounds
in the network theory~\eqref{eqn:short-path-type} in form of matrices
and to manipulate them using typed matrix multiplications.
For instance, we can combine the two outgoing edges of $A$ into a
single type
\begin{alignat}{1}
     (A \imp 5 \rcol \delm B) \;\wedge\; (A \imp 3 \rcol \delm C)
     \;\cong\;  A \imp (5, 3) \rcol \delm B \wedge \delm C
     \;\cong\;  [5;3] \rcol A \imp \delm B \wedge \delm C,
\label{eqn:a-column-vec}
\end{alignat}
where $[5;3]$ abbreviates the function $\lambda x.\, ((5,0),(3,0))$
interpreted as a \textit{column vector} of numbers. Dually, the two
incoming edges into node $D$ can be combined into a single type
\begin{alignat}{1}
     (B \imp 1 \rcol \delm D) \;\wedge\; (C \imp 4 \rcol \delm D)
    \;\cong\; [1,4] \rcol B \vee C \imp \delm D,
\label{eqn:a-row-vec}
\end{alignat}
where $[1,4]$ is the function $\lambda x.\, \;\mbox{case}\; x
\;\mbox{of}\; [0 \to (1,0), 1 \to (4,0)]$ thought of as a \textit{row}
vector. The type algebra, essentially~\eqref{eqn:sequential-comp}
and~\eqref{eqn:sum-comp-out}, proves that the conjunction of
both~\eqref{eqn:a-column-vec} and~\eqref{eqn:a-row-vec} implies the
matrix multiplication
\begin{alignat*}{1}
      & ([5;3] \rcol A \imp \delm B \wedge \delm C) \wedge ([1,4]
            \rcol B \vee C \imp \delm D)
      \;\preceq\; \mind(5 + 1, 3 + 4) \rcol A \imp \delm D 
      \;=\; [1,4]\cdot[5;3] \rcol A \imp \delm D 
\end{alignat*}
in \textit{min-plus} algebra.  More generally, for every sub-network
with source nodes $X_1, X_2, \ldots, X_m$ and sink nodes $Y_1, Y_2,
\ldots, Y_n$ we have an elementary type $D \rcol \vee_{i=1}^m X_i \imp
\wedge_{j=1}^n \delm Y_j$ describing the shortest path between any
source to any target, in which the scheduling bound $D \in
\sbound{(\vee_{i=1}^m X_i) \imp \otimes_{j=1}^n \delm Y_j}$ behaves
like a $n \times m$ matrix in min-plus algebra.  For instance, take
the decomposition of N into the edge sets $N_1 \df \{A \to B, A \to
C\}$, $N_2 \df \{B \to E, B \to D, C \to D, C \to F\}$ and $N_3 \df
\{D \to E, D \to F, E \to F\}$:
\begin{alignat*}{1}
    D(N_1) &= [5;3]  
           \;\rcol A \imp  (\delm B \wedge \delm C) \\
    D(N_2) &= [1;2;+\infty,4;+\infty;8]  
           \;\rcol (B \vee C) \imp 
                       (\delm D \wedge \delm E \wedge \delm F) \\
    D(N_3) &= [4,2,0]
           \;\rcol (D \vee E \vee F) \imp \delm F.
\end{alignat*}
The shortest path from $A$ to $F$ is then obtained by multiplying
these matrices
\begin{alignat*}{1}
     & [4,2,0] \cdot [1;2;+\infty,4;+\infty;8] \cdot [5;3]
         = [4,2,0] \cdot [6;7;11] = 9 \rcol A \imp \delm F 
\end{alignat*}
in min-plus-algebra. The type-theoretic approach facilitates a
compositional on-the-fly construction of the \textit{shortest path
  matrix}. The pure algebraic technique would combine all the
information in a global $6 \times 6$ network matrix $N \rcol (\vee_{X
  \in \vars} X) \imp (\wedge_{X \in \vars} \delm X)$ where $(N)_{XY} =
d < +\infty$ if there exists an edge $X \imp d \rcol Y$ in
$\phi_N$. Then, the shortest path matrix is $N^* = \mathit{Id} \wedge
N \wedge N^2 \wedge \cdots$, where $\mathit{Id}$ is the identity
matrix with $0$s in the diagonal and $+\infty$ everywhere else and
$\wedge$ is the operation of forming element-wise minimum, lifting the
logical operation $d_1 \rcol \delm X \wedge d_2 \rcol \delm X \cong
\mind(d_1, d_2) \rcol \delm X$ to matrices.  The entries in $N^*$ are
the shortest distances between any two nodes in the network.

This way of solving shortest paths is well-known, of course. But now
the behavioural typing permits us safely to play over- and
under-approximation games which are difficult to control in pure
algebra or graph theory without model-theoretic semantics.  Just to
give a simple example, suppose we wanted to derive a lower bound on
the shortest path. Such can be obtained by identifying some of the
control nodes, i.e., pretending we could jump between them on our path
to reach the destination. For instance, assuming $C \equiv B$, we find
that $\phi_N \wedge C \equiv B \preceq A \imp 7 \rcol \delm F$ is the
shortest distance. Since the conjunction $\phi_N \wedge C \equiv B$
specifies a subset of activations, the shortest distance between $A$
and $F$ relative to $\phi_N \wedge C \equiv B$ is a lower bound on the
shortest distance relative to $\phi_N$. It may be more efficient to
compute since the network $\phi_N \wedge C \equiv B$ only has $5$
different nodes rather than $6$ as with $\phi_N$.

\subsection{Task Scheduling}
\label{ex:task-sched-prob}

In yet another interpretation of network $N$ the nodes are tasks and
edges scheduling dependencies associated with upper bounds for task
completion.  Computing the worst-case completion time for the overall
schedule, sequential composition of edges corresponds to addition as
in the shortest path scenario Sec.~\ref{ex:short-path-prob} but
branching now involves maximum rather than the minimum. Again, this is
induced by the logical nature of the problem, the fact that the input
join now is conjunctive rather than disjunctive as before.  For
instance, task $D$ in Fig.~\ref{fig:sched-dep-grf} cannot start before
\textit{both} tasks $C$ \textit{and} $B$ have started with a set-up
delay of $4$ time units from the start of $C$ and $1$ unit from $B$.
Let us assume the task activation times are included in these set-up
delays. To model this type-theoretically we take the edges as the
atomic control variables, i.e., $\vars = \{AC, AB, CD, CF, BD, BE, DE,
DF, F\}$. Whenever $XY \in \sigma(i)$, for $i \in \length{\sigma}$,
this says that the edge $XY$ is ready, i.e., the source task $X$ is
completed and the start token has arrived at the corresponding control input
of target task $Y$.  The node $D$ establishes a logical-arithmetical
relationship between its input edges $CD$, $BD$ and its output edges
$DF$, $DE$, given by $CD \wedge BD \imp (4 \rcol \delm DF) \wedge (5 \rcol
\delm DE)$.  Overall,
\begin{alignat*}{1}
    \phi_N &\df
       (\true \imp 3 \rcol \delm AC \wedge 5 \rcol \delm AB)
       \wedge (AC \imp 4 \rcol \delm CD \wedge 8 \rcol \delm CF) \\
   &\quad \wedge (AB \imp 1 \rcol \delm BD \wedge 2 \rcol \delm BE) \wedge
       ((CD \wedge BD) \imp 4 \rcol \delm DF \wedge 5 \rcol \delm DE) \\
   &\quad \wedge (DE \wedge BE \imp 2 \rcol \delm EF)
          \wedge (CF \wedge DF \wedge EF \imp 0 \rcol \delm F).
\end{alignat*}
The critical path is the minimal $d$ such that $\phi_N \preceq d \rcol
\delm F$. It can be computed by linear programming involving
matrix multiplication in max-plus algebra using essentially the
laws~\eqref{eqn:sequential-comp} and~\eqref{eqn:and-abstr-out}.

\section{Examples II: Esterel-style Synchronous Multi-threading}
\label{sec:examples-2}

Like task scheduling in Sec.~\ref{ex:task-sched-prob}, the timing
analysis of Esterel programs~\cite{BoldtTvH07,MendlervHT09} involves
max-plus algebra, yet takes place in an entirely different fragment of
the type theory. Instead of implications 
$\zeta_1 \wedge \zeta_2 \imp \delm \xi_1
\wedge \delm \xi_2$ as in Sec.~\ref{ex:task-sched-prob} we employ
dependencies of the form $\zeta_1 \vee \zeta_2 \imp \delm \xi_1 \oplus
\delm \xi_2$, which are handled by~\eqref{eqn:sequential-comp}
and~\eqref{eqn:sum-abstr-in} rather than~\eqref{eqn:sequential-comp}
and~\eqref{eqn:and-abstr-out}. In addition, we use the tensor $\otimes$
for capturing multi-threaded parallelism.  Here we provide some
further theoretical background for the work reported
in~\cite{MendlervHT09}.

Esterel programs communicate via \textit{signals}, which are either
present or absent during one instant.  Signals are set present by the
\texttt{emit} statement and tested with the \texttt{present} test.
They are reset at the start of each instant.  Esterel statements can
be either combined in sequence (\texttt{;}) or in parallel
(\texttt{||}). The \texttt{loop} statement simply restarts its body
when it terminates.  All Esterel statements are considered
instantaneous, except for the \texttt{pause} statement, which pauses
for one instant, and derived statements like \texttt{halt} (=
\texttt{loop pause end}), which stops forever.  Esterel supports
multiple forms of preemption, \eg, via the \texttt{abort} statement,
which simply terminates its body when some trigger signal is present.
Abortion can be either weak or strong.  Weak abortion permits the
activation of its body in the instant the trigger signal becomes
active, strong abortion does not.  Both kinds of abortions can be
either immediate or delayed.  The immediate version already senses for
the trigger signal in the instant its body is entered, while the
delayed version ignores it during the first instant in which the abort
body is started.

\begin{figure}[ht]
\begin{minipage}[b]{0.5\textwidth}
    \subfloat[CKAG]{\label{fig:exampleckag}
      \includegraphics[width=.97\textwidth]{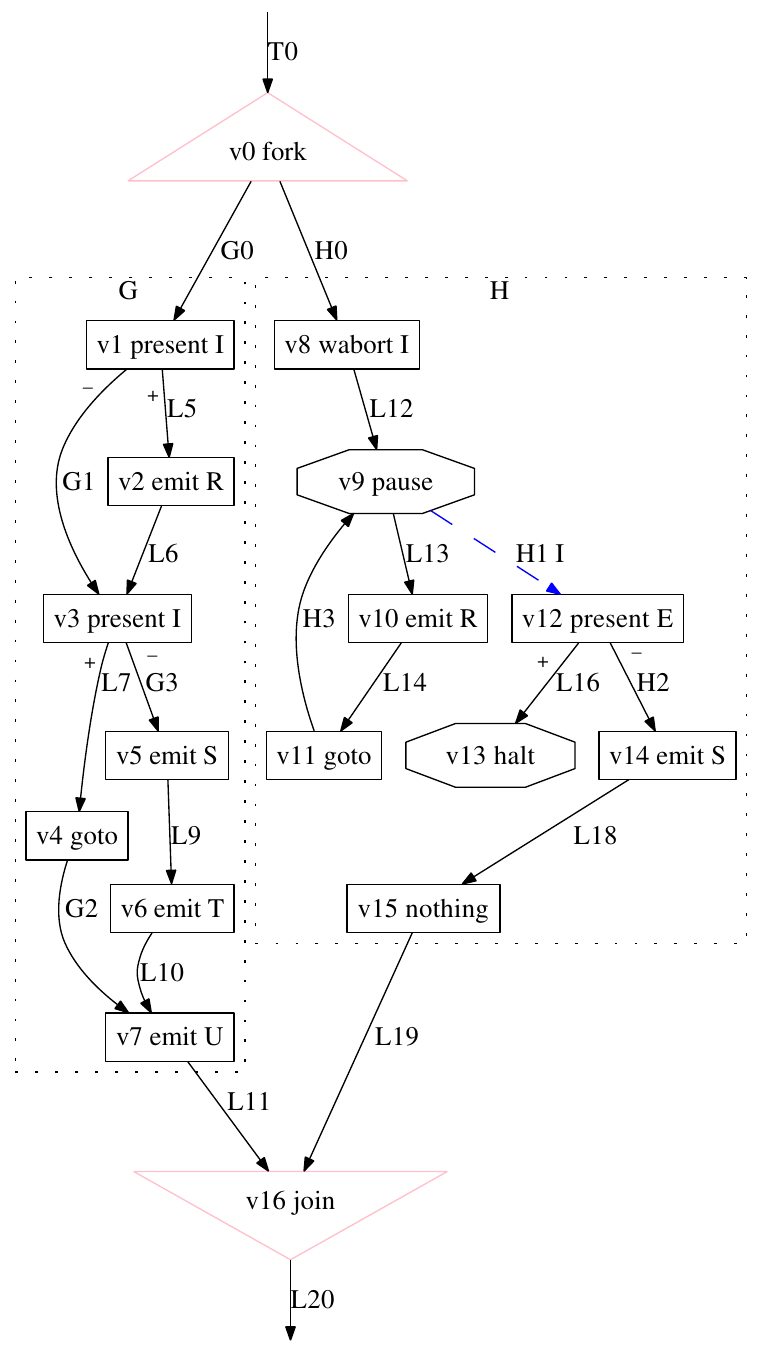}
    }
\end{minipage}
$\qquad$
\begin{minipage}[b]{0.4\textwidth}
\begin{SubFloat}{\label{fig:examplestrl}Esterel module $T$}
\begin{minipage}[b]{0.97\textwidth}
\begin{lstlisting}[language=Esterel,mathescape]
% module T
[ % thread G
  present I then emit R end present;
  present I else emit S; emit T end present;
  emit U;
$\mid\mid$
  % thread H
  weak abort
    loop
       pause;emit R
    end loop
  when immediate I;
  present E then halt end present;
  emit S; nothing;
]
\end{lstlisting}
\end{minipage}
\end{SubFloat}
\begin{center}
\begin{SubFloat}{\label{fig:examplekasm}KEP Assembler}
\begin{minipage}[b]{0.6\textwidth}
\begin{lstlisting}[language=KEP]
L01: T0: PAR 1,G0,1
L02:     PAR 1,H0,2
L03:     PARE A1

L04: G0: PRESENT I,G1
L05:     EMIT R
L06: G1: PRESENT I,G3
L07:     GOTO G2
L08: G3: EMIT S
L09:     EMIT T
L10: G2: EMIT U

L11: H0: WABORT I,H1
L12: H3: PAUSE
L13:     EMIT R
L14:     GOTO H3
L15: H1: PRESENT E,H2
L16:     HALT
L17: H2: EMIT S
L18:     NOTHING

L19: A1: JOIN
\end{lstlisting}
\end{minipage}
\end{SubFloat}
\end{center}    
\end{minipage}
  \caption{Esterel module $T$ (b) with 
    control-flow graph (a) and resulting KEP Assembler (c).}
  \label{fig:example}
\end{figure}

Consider the Esterel fragment in Figure~\ref{fig:examplestrl}. It
consists of two threads. The first thread $G$ emits signals
\texttt{R}, \texttt{S}, \texttt{T} depending on some input signal
\texttt{I}. In any case, it emits signal \texttt{U} and terminates
instantaneously. The thread $H$ continuously emits signal \texttt{R},
until signal \texttt{I} occurs.  Thereafter, it either halts, when
\texttt{E} is present, or emits \texttt{S} and terminates otherwise,
after having executed the skip statement \texttt{nothing}.

The \textit{concurrent KEP assembler graph}~\cite{LiBvH06} (CKAG, see
Fig.~\ref{fig:exampleckag}) captures the control flow, both standard
control and abortions, of an Esterel program.  The CKAG is derived
from the Esterel program by structural translation.  For a given CKAG,
the generation of assembly code for the Kiel Esterel Processor
(KEP)~\cite{LiBvH06,LivH10}, executing synchronous parallelism by
multi-threading, is straight-forward (see Fig.~\ref{fig:examplekasm}).

Let $\sigs$, $\labs$ and $\states$ be disjoint sets of (input or
output) \textit{signals}, control flow \textit{labels} and
synchronisation \textit{states}, respectively. For the Esterel module
in Fig.~\ref{fig:example} we have $\sigs = \{ I, E, R, S, T, U \}$,
$\labs = \{ L0, \ldots, L20, G0, \ldots, G3, H0, \ldots, H3 \}$. As
synchronisation states we use the names of the atomic delay nodes,
i.e., the \pause, \halt and \join nodes, $\states = \{ v_9, v_{13},
v_{16} \}$.  These describe the different state bits of the
synchronous automaton coded by the program block $T$. To distinguish the
cases of a thread starting from or ending in a given state $s \in
\states$ during an instant we use the modifiers $\start(s)$ and
$\wait(s)$. The former expresses that the thread is leaving from $s$
at the beginning of the instant and the latter that it enters and
terminates the instant in $s$. The set $\states^+ \df \{ \start(s),
\wait(s) \mid s \in \states \}$ collects these atomic statements. The
set of \textit{control variables}, specifying the atomic control
points of a program module, is the union $\vars = \sigs \cup \labs
\cup \states^+$.  All the controls $\start(s)$ are stable, i.e., we
may assume $\start(s) \oplus \neg\start(s)$. This is not true for
controls $\wait(s)$ which are switched on dynamically as the schedule
enters a delay node.

One possible activation of the Esterel module $T$ in
Fig.~\ref{fig:exampleckag} would be as follows.  Initially, control
variable $T0$ is set, so $\sigma(0) = \{ T0 \}$. Then the \PAR and
\PARE instructions making up the \fork node $v_0$ are executed in line
numbers L01, L02, L03 of Fig.~\ref{fig:examplekasm}, each taking one
instruction cycle (ic). The two \PAR instructions set up internal 
counters for thread
control, which does not change the set of events in the variables of
Fig.~\ref{fig:exampleckag}. Hence, $\sigma(1) = \sigma(2) = \{ T0 \}$.
After the \PARE both control variable $G0$, $H0$ become present
bringing threads $G$ and $H$ to life. This means $\sigma(3) = \{ T0,
G0, H0 \}$. The next instruction could be any of the two first
instructions of $G$ or $H$.  As it happens, the KEP Assembler
Fig.~\ref{fig:examplekasm} assigns higher priority to $H$ so that our
activation continues with \wabort (node $v_8$), i.e., $\sigma(4) =
\{T0, G0, H0, L12\}$. This brings up the \pause instruction $v_9$.
Now, depending on whether signal $I$ is present or not the activation
of \pause either moves to $v_{12}$ (weak immediate abort) or
terminates. Let us assume the latter, i.e., $\sigma(5) = \{T0, G0, H0,
L12, \wait(v_9) \}$, where thread $H$ is finished up for the instant
and has entered a wait state in node $v_9$. The activation continues
with the first instruction of $G$, the \present node $v_1$ at label
$G0$.  Since $I$ is assumed absent, its activation effects a jump to
label $G1$, i.e., $\sigma(6) = \{ T0, G0, H0, L12, \wait(v_9), G1 \}$.
Thereafter, we run sequentially through nodes $v_3$, $v_5$, $v_6$,
$v_7$ giving $\sigma(7) = \sigma(6) \cup \{ G3\}$, $\sigma(8) =
\sigma(7) \cup \{ L9 \}$ and $\sigma(9) = \sigma(8) \cup \{ L10 \}$.

Executing the final \emit instruction $v_7$ hits the \join at entry
$L11$, so that $\sigma(10)$ $=$ $\{ T0, G0, H0,$ $L12,$ $\wait(v_9), G1, G3,
L9, L10, L11 \}$.  Now both threads $G$ and $H$ are finished. While
$G$ is terminated and hands over to the main thread $T$ for good, $H$
is still pausing in $v_9$. It takes one activation step of the \join
node $v_{16}$ to detect this and to terminate the synchronous instant
of $T$ with the final event $\sigma(11)$ $=$ $\{ T0, G0, H0, L12,
\wait(v_9), G1, G3, L9, L10, L11, \wait(v_{16}) \}$.  
Over\-all, we get an acti\-va\-tion of the
outer-most main thread of $T$, 
$\sigma = \sigma(0), \ldots, \sigma(11)$, starting from program label $T0$
consisting of $12$ ics in total.  In the next logical
instant when $T$ is resumed in $v_{16}$ and $v_9$, with initial event
$\sigma(0) = \{\start(v_9), \start(v_{16})\}$, and thread $H$
eventually comes out at control point $L19$ (if signal $I$ is present
and $E$ absent), then executing the \join $v_{16}$ will bring us to
control point $L20$ and out of $T$ instantaneously.

Activation sequences starting in control label $T0$ and ending in $L20$ are
called \textit{through paths}, those starting in $T0$ and pausing in a
synchronisation state $\wait(s)$, $s \in \{ v_9, v_{13}, v_{16} \}$,
are \textit{sink paths}; \textit{source paths} begin in a state
$\start(s)$ and end in $L20$, while \textit{internal paths} begin in a
state and end in a state.

\paragraph{Esterel IO-Interface Types.}

Our normal form interfaces to describe Esterel-KEP modules are of the
form $\theta = \phi \imp \psi$, with \emph{input control} $\phi =
\bigvee_{i=1}^m \zeta_i$ and \emph{output control} $\psi =
\bigoplus_{k=1}^n \delm \xi_k$ where the $\zeta_{i}$ and $\xi_{k}$ are
pure types.  The former $\phi$ captures all the possible ways in which
a program module (or any other fragment) of type $\theta$ can be
started within an instant and the latter $\psi$ sums up the ways in
which it can be exited during the instant.  Intuitively, $\Sigma
\models \theta$ says that whenever the schedule $\Sigma$ enters the
fragment through one of the input controls $\zeta_i$ then within some
bounded number of ics it is guaranteed to exit through one of the
output controls $\xi_k$.  The disjunction $\vee$ in the input control
$\phi$ models the \emph{external} non-determinism resolved by the
environment which determines how a program block is started.  On the
output side $\psi$, the selection of which exit $\xi_k$ is taken is
expressed by $\oplus$ since it is an \emph{internal} choice which is
dynamically resolved during each activation. Each delay operator
$\delm$ stands for a possibly different delay depending on which
output $\xi_k$ is taken. Contrast this with an output control such as
$\psi = \delm(\bigoplus_{k=1}^n \xi_k)$ which only specifies one bound
for all exits $\xi_k$.  An interface bound $T \in \sbound{\phi \imp
  \psi}$ can be understood as a $n \times m$ shaped timing matrix
relative to the Boolean controls $\zeta_i$ and $\xi_k$ serving as
``base'' vectors.  The logical conjunction of these interfaces in a
fixed set of such base controls corresponds to matrix multiplications
in max-plus algebra. Furthermore, using logical reasoning on base
controls $\zeta_i$, $\xi_j$ we can massage the semantics of timing
matrices very much like we do with base transformations in ordinary
linear algebra. Two important operations on IO-interfaces are matrix
multiplication and the Kronecker product which in our scheduling
algebra are now strongly typed and thus receive semantic meaning in
logical spaces.

\paragraph{Transient and Sequential Submodules $G$ and $H$.}
\label{sec:submodule-GH}

A full and exact WCRT specification encapsulating the synchronous
block $G$ as a component would require mention of program labels $G1$,
$G3$, $G2$ which are accessible from outside for jump statements.
Therefore, the interface type for single-threaded scheduling of $G$
would be $ [6, 4, 3, 1] \rcol G0 \vee G1 \vee G3 \vee G2 \imp \delm
L11$.  This is still not the exact description of $G$ since it neither
expresses the dependency of the WCRT on signal $I$, nor the emissions
of $R$, $S$, $T$, $U$. For instance, if $I$ is present then all
threads must take control edges L5 and L7 rather than G1 or G3 which
are blocked. If $I$ is absent then both G1 and G3 must be taken
instead. As a result the longest path 
$v_1 + v_2 + v_3 + v_5 + v_6 + v_7$
with delay $6$ is not executable. To capture this, we consider signal
$I$ as another control input and refine the WCRT interface type of G:
\begin{equation}
  [5, 5, 3, 4, 3, 1] 
    \rcol (G0 \wedge I) \vee (G0 \wedge \neg I) \vee 
             (G1 \wedge I) \vee (G1 \wedge \neg I) 
               \vee G3 \vee G2 \imp \delm L11.
\label{eqn:intfc}
\end{equation}
The inclusion of signal $I$ in the interface has now resulted in the
distinction of two \textit{different} delay values $3$ and $4$ for $G1
\imp \delm L11$ depending on whether $I$ is present or absent. On the
other hand, $G0$, split into controls $G0 \wedge I$ and $G0 \wedge \neg
I$, produces the same delay of $5$ ics in both cases, which is a
decrease of WCRT compared to $[6] \rcol G0 \imp \delm L11$ from above.
Assuming that input signal $I$ is causally stable, i.e., $I \oplus
\neg I \cong \true$, it is possible to optimise the interface without
losing precision: since $(G0 \wedge I) \oplus (G0 \wedge \neg I) \cong
G0 \wedge (I \oplus \neg I) \cong G0 \wedge \true \cong G0$ the column
vector $[0; 0] \rcol G0 \imp \delm (G0 \wedge I) \oplus \delm (G0
\wedge \neg I)$ is sound and can be used to compress the two entries
of value $5$ in \eqref{eqn:intfc} into a single value $5 = \maxd(5,
5)$ giving $[5, 3, 4, 3, 1] \rcol G0 \vee (G1 \wedge I) \vee (G1
\wedge \neg I) \vee G3 \vee G2 \imp \delm L11$.  In the same vein, but
this time without referring to stability, we could further bundle $G1
\wedge I$ and $G3$ into a single control with the single delay $[3]
\rcol (G1 \wedge I) \oplus G3 \imp \delm L11$ at the same level of
precision. This finally yields $[5, 3, 4, 1] \rcol G0 \vee ((G1 \wedge
I) \oplus G3) \vee (G1 \wedge \neg I) \vee G2 \imp \delm L11$.  Still,
if we only ever intend to use $G$ as an encapsulated block with entry
$G0$ and exit $L11$ the following typing is sufficient:
\begin{equation}
   [5] \rcol G0 \imp \delm L11.
\label{eqn:G-interface}
\end{equation}

Now we take a look at the sequential control flow which starts and
terminates in \pause and \halt nodes. Consider the sub-module $H$ from
Fig.~\ref{fig:exampleckag} consisting of nodes $v_8$--$v_{15}$. Nodes
\wabort, \emit, \goto, \present, \nothing are transient and specified
as before for $G$. But now the instantaneous paths are broken by the
delay nodes $v_9$ and $v_{13}$.

First, consider the \pause node $v_9$. It can be entered by two
controls, line number L12 and program label H3, and left via two
exits, a non-instantaneous edge L13 and an instantaneous exit H1 (weak
abortion). When a control thread enters $v_9$ then either it
terminates the current instant inside the node or leaves through the
weak abort H1 (data-dependent, if signal $I$ is present) continuing
the current reaction, instantaneously.  A thread entering $v_9$ never
exits through L13 in the same instant. On the other hand, if a thread
is started (resumed) from inside the \pause node $v_9$ then control
can only exit through L13.  This suggests to specify the \pause node
as follows:
\begin{eqnarray}
     [1;1,1;1]
      &\rcol& H3 \vee L12 \imp \delm H1 \oplus \delm \wait(v_9)
 \label{eqn:pause-type-a} \\ \mbox{}
    [1] &\rcol& \start(v_9) \imp \delm L13.
 \label{eqn:pause-type-b}
\end{eqnarray}
The interface~\eqref{eqn:pause-type-a} says that if \pause is entered
through H3 or L12 it can be left through H1 or terminate ($\wait$)
inside the \pause. In all cases activation takes $1$ instruction
cycle.  Since there are no differences in the delays we could bundle
the controls $H3$, $L12$ and compress the matrix~\eqref{eqn:pause-type-a}
as $[1] \rcol H3 \oplus
L12 \imp \delm (H1 \oplus \wait(v_9))$ without losing information. 
We could also record the
dependency of control on signal $I$, with the more precise interface
$[1;-\infty,-\infty,1] \rcol ((H3 \oplus L12) \wedge I) \vee ((H3
\oplus L12) \wedge \neg I) \imp \delm H1 \oplus \delm \wait(v_9)$.
This separates the threads which must stop inside the pause from those
which must leave via H1 due to a weak immediate abort on signal $I$.
The specification \eqref{eqn:pause-type-b} accounts for threads
starting in the \pause which must necessarily pass control to L13
within one instruction cycle.

The \halt node $v_{13}$ in Fig.~\ref{fig:exampleckag} is not only a
sink for control threads entering through L16 but it also has an
internal path of length $1$ (which is repeated at every instant). It
is specified by the interface $ [1, 1] \rcol (\start(v_{13}) \vee L16)
\imp \delm \wait(v_{13})$.  By composition from the WCRT interfaces of
nodes $v_{12}$--$v_{15}$ using matrix multiplications in max-plus
algebra we get
\begin{alignat}{1}
  H &= 
    [5;4, 7;6]
    \rcol H0 \vee \start(H) \imp \delm L19 \oplus \delm \wait(H)
\label{eqn:H-interface}
\end{alignat}
recording the lengths of the longest through path 
$v_8 + v_9 + v_{12} + v_{14} + v_{15}$, sink path $v_8 + v_9 + v_{12} + v_{13}$,  
source path $v_9 + v_{10} + v_{11} + v_{9} + v_{12} + v_{14} + v_{15}$ and 
internal path $v_9 + v_{10} + v_{11} + v_{9} + v_{12} + v_{13}$.

\paragraph{Multi-threading Composition: Fork and Join.}
\label{sec:fork-join}

Finally, consider the two blocks $G$ and $H$ as they are combined
inside the Esterel module $T$ (Fig.~\ref{fig:exampleckag}) and
synchronised by \fork and \join nodes $v_0$ and $v_{16}$.  The main
thread starts $G$ and $H$ in their initial controls, i.e., by
activating $G0 \wedge H0$. Then, the executions of $G$ and $H$ are
interleaved, depending on the priorities assigned by the compiler
about which we shall make no assumptions. Child thread $G$ can only
run through its instantaneous path until it reaches L11 where it is
stopped by the \join. The sequential block $H$ has two options: It can
take its instantaneous through path stopping at $L19$ or it pauses in
one of its delay nodes.  In the former case we have reached $L11
\wedge L19$, where the synchronising \join takes over letting the main
thread continue by instantaneously activating L20 within the same
instant. In the latter case we have activated $L11 \wedge \wait(H)$
where the synchronous instant is finished and the combined system
pauses.  Activation is resumed in the next instant from $L11 \wedge
\start(H)$, while $G$ is still inactive and waiting at $L11$.  Child
thread $H$ may either leave instantaneously through $L19$, giving $L11
\wedge L19$ overall, or once more pause internally, leading again to
$L11 \wedge \wait(H)$.

This synchronous composition is obtained by the Kronecker product $GH
\df G' \otimes H'$ where $G'$ and $H'$ are the stand-alone interfaces
of $G$~\eqref{eqn:G-interface} and $H$~\eqref{eqn:H-interface}
instrumented for the synchronisation:
\begin{alignat*}{1}
G' &= \textit{Sync}_1 \wedge [5, 0] \rcol   
      G0 \vee L11 \imp \delm L11 \\
H' &= \textit{Sync}_2 \wedge [5;4, 7;6] \rcol 
      H0 \vee \start(H) \imp 
           \delm L19 \oplus \delm\wait(H).
\end{alignat*}
$G$ is extended by the additional input control $L11$ and trivial path
$[0] \rcol L11 \imp \delm L11$ to let $G$ start an instant from $L11$
when $H$ is pausing.  The conjunct $\textit{Sync}_1 \df \neg L11$
expresses the synchronisation whereby $G$ finishes once it reaches
$L11$.  Similarly, the conjunct $\textit{Sync}_2 \df \neg(L19 \oplus
\wait(H))$ added to the interface~\eqref{eqn:H-interface} stops $H$
from continuing its activation instant past $L11$ or $\wait(H)$.  The
Kronecker product $G' \otimes H'$ now generates all possible
interleaving of activations specified by type $G'$ with those from
type $H'$:
\begin{alignat*}{1}
  G' \otimes H' &\preceq
    [5, 0] \otimes [5;4, 7;6] \;=\;
     [5 \cdot [5;4, 7;6], 0 \cdot [5;4, 7;6]] \;=\; [10;9, 12;11, 5;4, 7;6] \\
   &\rcol  (G0 \wedge H0) \vee
            (G0 \wedge \start(H)) \vee 
             (L11 \wedge H0) \vee 
               (L11 \wedge \start(H))
   \imp \delm(L11 \wedge L19) \oplus \delm(L11 \wedge \wait(H)).
\end{alignat*}
In the synchronised composition $GH$ we are only interested in the
(surface) paths initiated by $G0 \wedge H0$ and the (depth) paths
activated by the combination $L11 \wedge \start(H)$. All other paths
cannot be activated inside the \fork and \join context. Thus, we drop
these column vectors and only continue with
\begin{eqnarray*}
  GH &=& [10;9, 12;11, 5;4, 7;6] \cdot 
           [0;-\infty;-\infty;-\infty, -\infty;-\infty;-\infty,0] 
             \;=\; [10;9, 7;6] \\
     && \rcol (G0 \wedge H0) \vee (L11 \wedge \start(H)) \imp 
        \delm(L11 \wedge L19) \oplus \delm(L11 \wedge \wait(H)).
\end{eqnarray*}
This models the concurrent composition of $G$ and $H$ but not yet the
interface of the composite block $T$ with \fork and \join as depicted in
Fig.~\ref{fig:exampleckag}. These are additional components specified
as
\begin{xalignat*}{1}
  \mbox{\join} = 
  & [1;-\infty, -\infty;1] 
     \rcol (L11 \wedge L19) \vee (L11 \wedge \wait(H)) \imp 
                 \delm L20 \oplus \delm\wait(T) \\
  \mbox{\fork} =
  & [3;-\infty, -\infty;0] 
     \rcol T0 \vee \start(T) \imp 
        \delm(G0 \wedge H0) \oplus \delm(L11 \wedge \start(H))
\end{xalignat*}
with new state controls $\wait(T)$ and $\start(T)$ for module $T$. 
The \JOIN instruction in line 19 of Fig.~\ref{fig:examplekasm}
is always executed upon termination of both threads from $G$ and $H$
inside $T$ and the associated activation time of one ic is accounted
for in the \join interface above. Specifically, this is a through path
$[1] \rcol (L11 \wedge L19) \imp \delm L20$ and source path $[1] \rcol
L11 \wedge \wait(H) \imp \delm\wait(T)$.  The entry $[3] \rcol T0 \imp
\delm(G0 \wedge H0)$ of $\fork$ includes the ics for two \PAR, one
\PARE from lines 1-3 of Fig.~\ref{fig:examplekasm}.  
Adding \fork and \join on the input and output side then obtains
\begin{alignat*}{1}
  T &= [1;-\infty, -\infty;1] \cdot 
        [10;9, 7;6] \cdot 
        [3;-\infty, -\infty;0] 
    \;= [14;13, 8;7] 
           \;\rcol\; T0 \vee \start(T) \imp \delm L20 \oplus \delm\wait(T)
\end{alignat*}
for the composite module $T$.  Indeed, the longest through path is
exemplified by the sequence of nodes $v_0 (3) + \{ v_1 + v_2 + v_3 +
v_4 + v_7 \}_G (5) + \{ v_8 + v_9 + v_{12} + v_{14} + v_{15} \}_H (5)
+ v_{16} (1) = 14$. A longest sink path is $v_0 (3) + \{ v_1 + v_2 +
v_3 + v_4 + v_7 \}_G (5) + \{ v_8 + v_9 + v_{12} + v_{13} \}_H (4) +
v_{16} (1) = 13$. As a maximal source path we could take $\{ \}_G (0)
+ \{v_9 + v_{10} + v_{11} + v_9 + v_{12} + v_{14} + v_{15} \}_H (7) +
v_{16} (1) = 8$ and as a possible longest internal path $\{ \}_G (0) +
\{ v_9 + v_{10} + v_{11} + v_9 + v_{12} + v_{13} \}_H (6) + v_{16} (1)
= 7$.

In specific WCRT algorithms such as the one of~\cite{BoldtTvH07} many
of the matrix multiplications shown above are executed efficiently in
the combinatorics of traversing the program's control flow graph
forming maximum and additions as we go along. This is possible only so
far as control flow dependencies are represented explicitly in the
graph. In general, with data-dependencies, this may be an exponential
problem so that symbolic techniques for modular analyses are needed.
Our logical interface algebra can be used to keep track of the
semantic meaning of WCRT data. Even without data-dependencies, the
WCRT interfaces presented here give rise to a depth-first search
algorithm~\cite{MendlervHT09} which is already more precise than the
one presented in~\cite{BoldtTvH07}.

\section{Related Work}
\label{sec:conclusion}

Most interface models in synchronous programming are restricted to
causality issues, \ie, dependency analysis without considering
quantitative time. Moreover, the granularity of dependency is limited.
E.g., the modules of Andr{\'e} \etal~\cite{AndreBPRV97} do not permit
instantaneous interaction.  Such a model is not suitable for
compositional, intra-instant, scheduling analysis.  Hainque
\etal~\cite{HainquePBN99} use a topological abstraction of the
underlying circuit graphs (or syntactic structure of Boolean
equations) to derive a fairly rigid component dependency model.  A
component is assumed executable iff \textit{all} of its inputs are
available; after component execution \textit{all} of its outputs
become defined. This is fine for concurrent execution but too
restricted to model single- or multi-threaded execution
compositionally.  The interface model also does not cover data
dependencies and thus cannot deal with dynamic schedules. It also does
not support quantitative resource information, either.

The causality interfaces of Lee \etal~\cite{LeeZZ05} are much more
flexible.  These are functions associating with every pair of input
and output ports an element of a \textit{dependency domain} $D$, which
expresses if and how an output depends on some input.  Causality
analysis is then performed by multiplication on the global system
matrix.  Using an appropriate dioid structure $D$, one can perform the
analyses of Hainque et.\ al.~\cite{HainquePBN99} as well as restricted
forms of WCRT. Lee's interfaces presuppose a fixed static distinction
between inputs and outputs and cannot express the difference between
an output depending on the joint presence of several values as opposed
to depending with each input individually.  Similarly, there is no
coupling of outputs, \eg, that two outputs always occur together at
``the same time.'' Thus, they do not support full AND- and OR-type
synchronisation dependencies for representing multi-threading and
multi-processing.  Also, the model does not include data dependency.
The work reported here can be seen as an extension of~\cite{LeeZZ05}
to include such features. In particular, note that our scheduling
interfaces can also be used in situations where linear algebra
is not applicable, as in the case of network flow problems.

Recent works~\cite{WT:EMSOFT05,HM:RTAS06} combining network
calculus~\cite{BaccelliCOQ:SyncNLin,LeBTh:network-calculus} with
real-time interfaces are concerned with the compositional modelling of
regular execution patterns.  Existing interface
theories~\cite{LeeZZ05,WT:EMSOFT05,HM:RTAS06}, which aim at the
verification of resource constraints for real-time scheduling, handle
timing properties such as task execution latency, arrival rates,
resource utilisation, throughput, accumulated cost of context
switches, and so on. The dependency on data and control flow
is largely abstracted. For instance, since the task sequences of
Henzinger and Matic~\cite{HM:RTAS06} are independent of each other,
their interfaces do not model concurrent forking and joining of
threads.  The causality expressible there is even more restricted than
that by Lee \etal~\cite{LeeZZ05} in that it permits only one-to-one
associations of inputs with outputs.  The interfaces of Wandeler and
Thiele~\cite{WT:EMSOFT05} for modular performance analysis in
real-time calculus are like those of Henzinger and
Matic~\cite{HM:RTAS06} but without sequential composition of tasks and
thus do not model control flow. On the other hand, the 
approaches~\cite{WT:EMSOFT05,HM:RTAS06} can describe continuous and 
higher-level stochastic properties which our interface types cannot.

AND- and OR-type synchronisation dependencies are important for
synchronous programming since reachability of control nodes in general
depends both conjunctively and disjunctively on the presence of data.
Also, control branching may be conjunctive (as in multi-threading or
concurrent execution) or disjunctive (as in single-threaded code).
Moreover, execution may depend on the absence of data (negative
triggering conditions), which makes compositional modelling rather a
delicate matter in the presence of logical feedback loops.  This
severely limits the applicability of existing interface models.  The
assume-guarantee style specification~\cite{WT:EMSOFT05,HM:RTAS06} does
not address causality issues arising from feedback and negative
triggering conditions. The interface automata of Alfaro, Henzinger,
Lee, Xiong~\cite{deAH:interface-automata,LeeX:SysTypes} model
synchronous macro-states and assume that all stabilisation processes
(sequences of micro-states) can be abstracted into atomic interaction
labels. The introduction of \textit{transient
  states}~\cite{LeeX:BehTypes} alleviates this, but the focus is still
on regular (scheduling) behaviour. The situation is different,
however, for cyclic systems, in which causality information is needed.
Our interface algebra is semantically sound with respect to feedback
and indeed supports causality analysis as a special case: A signal $A$
is \textit{causal} if $\delm A \oplus \neg A$ can be derived in the
type theory of a module.  Because of the complications arising from
causality issues, there is currently no robust component model for
synchronous programming.  We believe that the interface types
introduced in this paper, cover new ground towards such a theory.

Finally, note that our algebra is not intended as a general purpose
interface model such as, e.g., the \textit{relational interfaces} of
Tripakis \etal~\cite{TripLickHenzLee:RelInterfaces}.  While these
relational interfaces permit contracts in first-order
logic between inputs and outputs, our interfaces only describe
propositional relations. Therefore, our algebra cannot describe the
full functional behaviour of data processing (other than by coding it
into finite Booleans).  Our interfaces are logically restricted to
express monotonic scheduling processes and the resource consumption
inside synchronous instants. Because we use an intuitionistic
realisability semantics (Curry-Howard) we obtain enough expressiveness
to deal with causality problems and upper-bound scheduling costs.  The
interface algebra does not aim to cover behavioural aspects of
sequences of instants such as in approaches based on temporal logics
or the timed interfaces of Alfaro, Henzinger and
Stoelinga~\cite{deAlfHenzStoe:TimedInterfaces}, which build on timed
automata. The scheduling problem addressed here is a simpler problem
in the sense that it arises afresh \textit{within} each synchronous
step and does not need to carry (e.g., timing) constraints
\textit{across} steps. However, note that our algebra can fully
capture finite-state sequential transition functions in the standard
way by duplicating propositional state variables $s$ using $\start(s)$
and $\wait(s)$ as seen in Sec.~\ref{sec:examples-2}. An
\emph{inter-instant} transition (instantaneous, no clock tick) between
$s_1$ and $s_2$ is given by the implication $\start(s_1) \imp \delm
\wait(s_2)$ while the \emph{intra-instant} transition (sequential,
upon clock tick) is the weak implication $\neg\wait(s_1) \oplus
\start(s_2)$. In this way, we can derive exact state-dependent
worst-case bounds across all reachable states of a finite state
behaviour.

The scheduling algebra in this paper
extends~\cite{mvm00:IGPL} in that it not only captures
concurrent execution (as in combinational circuits) but also includes
the tensor $\otimes$ for multi-threading. More subtly,
while~\cite{mvm00:IGPL} is restricted to properties of activation
sequences stable under the \emph{suffix} preordering, here we consider
the much richer lattice of \emph{arbitrary} sub-sequences. This paper
introduces the theory behind~\cite{MendlervHT09} which reported on the
application to WCRT analysis for Esterel and also
provides more detailed information on the modelling in
Sec.~\ref{sec:examples-2}.

\paragraph{Acknowledgements.}

The author would like to thank the anonymous reviewers for their
suggestions to improve the presentation.


\bibliographystyle{eptcs}


\end{document}